\documentclass[twocolumn]{aastex63}
\usepackage{amsmath}
\shorttitle{AT2020hur}
\shortauthors{Li et al. 2021}

\begin{document}

\title{AT2020hur: A Possible Optical Counterpart of FRB 180916B}

\author[0000-0002-8391-5980]{Long Li}
\affiliation{School of Astronomy and Space Science, Nanjing University, Nanjing 210023, China}

\author[0000-0003-0855-3649]{Qiao-Chu Li}
\affiliation{School of Astronomy and Space Science, Nanjing University, Nanjing 210023, China}

\author[0000-0002-1766-6947]{Shu-Qing Zhong}
\affiliation{Department of Astronomy, School of Physical Sciences, University of Science and Technology of China, Hefei 230026, China}

\author{Jie Xia}
\affiliation{Purple Mountain Observatory, Chinese Academy of Sciences, Nanjing 210008, China}
\affiliation{School of Astronomy and Space Science, University of Science and Technology of China, Hefei 230026, China}

\author[0000-0003-1946-086X]{Lang Xie}
\affiliation{Purple Mountain Observatory, Chinese Academy of Sciences, Nanjing 210008, China}
\affiliation{School of Astronomy and Space Science, University of Science and Technology of China, Hefei 230026, China}

\author[0000-0003-4157-7714]{Fa-Yin Wang}
\affiliation{School of Astronomy and Space Science, Nanjing University, Nanjing 210023, China}
\affiliation{Key Laboratory of Modern Astronomy and Astrophysics (Nanjing University), Ministry of Education, China}

\author[0000-0002-7835-8585]{Zi-Gao Dai}
\affiliation{Department of Astronomy, School of Physical Sciences, University of Science and Technology of China, Hefei 230026, China; daizg@ustc.edu.cn}
\affiliation{School of Astronomy and Space Science, Nanjing University, Nanjing 210023, China}

\begin{abstract}

The physical origin of fast radio bursts (FRBs) remains unclear. Finding multiwavelength counterparts of FRBs can be a breakthrough in understanding their nature. In this work, we perform a systematic search for astronomical transients (ATs) whose positions are consistent with FRBs. We find an unclassified optical transient AT2020hur ($\alpha=01^{\mathrm{h}} 58^{\mathrm{m}} 00.750^{\mathrm{s}} \pm 1$ arcsec, $\delta=65^{\circ} 43^{\prime} 00.30^{\prime \prime} \pm 1$ arcsec; \citealt{Lipunov2020}) that is spatially coincident with the repeating FRB 180916B ($\alpha=01^{\mathrm{h}} 58^{\mathrm{m}} 00.7502^{\mathrm{s}} \pm 2.3$ mas, $\delta=65^{\circ} 43^{\prime} 00.3152^{\prime \prime} \pm 2.3$ mas; \citealt{Marcote2020}). The chance possibility for the AT2020hur-FRB 180916B association is about 0.04\%, which corresponds to a significance of $3.5\sigma$.
We develop a giant flare afterglow model to fit AT2020hur. Although the giant flare afterglow model can interpret the observations of AT2020hur, the derived kinetic energy of such a GF is at least three orders of magnitude larger than that of the typical GF,
and there is a lot of fine tuning and coincidences required for this model. Another possible explanation is that AT2020hur might consist of two or more optical flares originating from the FRB source, e.g. fast optical bursts produced by the inverse Compton scattering of FRB emission. Besides, AT2020hur is located in one of the activity windows of FRB 180916B, which is an independent support for the association. This coincidence may be due to the reason that the optical counterpart is subject to the same periodic modulation as FRB 180916B, as implied by the prompt FRB counterparts. Future simultaneous observations of FRBs and their optical counterparts may help reveal their physical origin.
\end{abstract}


\section{Introduction} \label{sec:intro}

Fast radio bursts (FRBs) are transient radio pulses of millisecond duration with extremely high brightness temperatures (e.g., \citealt{Lorimer2007,Thornton2013,Cordes2019,Petroff2019}), and their physical origin remains a puzzle (see \citealt{Katz2018,Cordes2019,Petroff2019,Platts2019,ZhangB2020,Xiao2021}). They usually have a large dispersion measure (DM) exceeding the Galactic contribution, which suggests that most FRBs are extragalactic (e.g., \citealt{Lorimer2007,Thornton2013}). FRB 200428 is the first discovered Galatic FRB associated with an X-ray burst from the Galactic magnetar SGR 1935+2154 \citep{Bochenek2020,CHIME2020b,Mereghetti2020,Li2021,Ridnaia2021,Tavani2021}. The association between FRB 200428 and the X-ray burst has been discussed based on various models (e.g., \citealt{Dai2020b,Lu2020,Lyutikov2020b,Margalit2020}). The discovery of FRB 200428 suggests that at least some FRBs are of magnetar origin.

Most FRBs appear to be one-off, and only about twenty FRBs are now known to repeat. In general, repeating FRBs seem to repeat in an irregular way. However, a period of 16.35 days with a $\sim5$-day active window has been reported for FRB 180916B \citep{CHIME2020a}. Follow-up broad-band radio observations of FRB 180916B updated the period to 16.29 days with a 6.1-day active window \citep{Aggarwal2020,Marthi2020,Pastor2020,Sand2020}. Several models were proposed to explain the periodic activity of FRB 180916B \citep[e.g.,][]{Dai2020a,Ioka2020,Levin2020,Lyutikov2020a,Tong2020,Yang2020a,Zanazzi2020,geng21,liqc21,Wei2022}. Besides, a possible period of $\sim157$ days for FRB 121102 was also suggested by \cite{Rajwade2020}.
The luminosity distance of FRB 180916B is about an order of magnitude lower than that of FRB 121102 \citep{tend17, Marcote2020}.
Thus it will be easier for multiwavelength counterparts of FRB 180916B to be detected.
For FRB 180916B, there have been a lot of multiwavelength follow-up, simultaneous or during active-phase observations in the optical, X-ray and $\gamma$-ray frequency bands, but with no transient counterpart \citep[e.g.,][]{and20,cas20,sch20,Tavani2020b, Kilpatrick2021}.

So far, there are two repeating FRBs (FRB 121102 and FRB 190520B) accompanied by compact persistent radio emissions\footnote{FRB 20201124A was also associated with a radio counterpart, however, it was suggested that the radio counterpart comes from the star formation of its host galaxy \citep{Ravi2021}.} with the specific luminosities of the order of $\sim10^{29}\ \rm erg\ s^{-1}\ Hz^{-1}$ at GHz frequencies \citep{Chatterjee2017,Niu2021}. It is generally believed that the persistent radio emission of FRB 121102 is related to a young magnetar wind nebula \citep{Murase2016,Beloborodov2017,Metzger2017,Margalit2018, liqc20}.
Notably, there is a radio emission coincident with the superluminous supernova (SLSN) PTF10hgi, with the luminosity and frequency consistent with persistent radio emission of FRB 121102 \citep{eft19}. This implies that there may exist some connections between FRBs and other transient sources. However, with the late-time radio observations to some type-I SLSNe and gamma-ray burst remnants, there is still no FRBs detected \citep{law19, Men2019}.

Except for the X-ray burst associated with FRB 200428 and the persistent radio emissions associated with FRB 121102 and FRB 190520B, there are no confirmed multiwavelength counterparts that are associated with other FRBs (e.g. \citealt{Petroff2015,Callister2016,Gao2017,Zhang2017,MAGIC2018,Tingay2019,ZhangG2020}).
Many FRB models predict various multiwavelength counterparts, which can be classified into the following categories: (1) prompt FRB multiwavelength counterparts (e.g. \citealt{Metzger2019,Yang2019b,Beloborodov2020,Chen2020,Dai2020b}); (2) FRB multiwavelength afterglows (e.g. \citealt{Yi2014}); (3) counterparts arising from the circumburst environments (e.g. the surrounding nebula emission; various ``cosmic comb" associated with FRBs, as suggested by \citealt{ZhangB2017}; optical counterparts from FRBs heating companion stars in close binary systems, as suggested by \citealt{Yang2021}); (4) counterparts produced by the progenitor systems of FRBs (e.g. \citealt{ZhangB2014,Murase2016,Metzger2017,Wang2020a,Wang2020b}). The prompt multiwavelength counterparts, which usually have short durations ($\lesssim$ 100 seconds), are summarised and discussed by \cite{Chen2020}, including theoretical predicted counterparts from \cite{Metzger2019} model and \cite{Beloborodov2020} model, magnetar giant flares as FRB counterparts, and fast optical bursts associated with FRBs suggested by \cite{Yang2019b}. The other types of FRB counterparts usually have long duration ($\gtrsim$ 100 seconds), e.g., long-lasting multiwavelength FRB afterglows, persistent radio emission possible associated with the surrounding nebula, or supernovae that may have originated in the FRB progenitor systems.

The non-detections of multiwavelength counterparts of FRBs may be due to the following reasons (e.g., \citealt{Wang2020b}): (1) the fluxes of the multiwavelength counterparts are too faint to be detected, such as the multiwavelength afterglows of FRBs \citep{Yi2014}; (2) the durations of the multiwavelength counterparts are too short relative to the time resolution of a detector, e.g. the fast optical burst produced by one-zone inverse Compton scattering process could be as short as the duration of the FRB itself \citep{Yang2019b}; (3) the delay times between FRBs and their multiwavelength counterparts are too long compared with the observation time, such as supernovae and gamma-ray bursts that may be produced in the FRB progenitor systems may have a very long time delay with respect to the FRB itself. Many efforts have been made to search the multiwavelength counterparts of FRBs (e.g., \citealt{Bannister2012,Palaniswamy2014,DeLaunay2016,Scholz2016,Yamasaki2016,Xi2017,Zhang2017,Cunningham2019,Guidorzi2019,Men2019,Yang2019a,Tavani2020b,Wang2020b}). However, there are no confirmed results so far.

In this paper, we perform a systematic search for astronomical transients (ATs) that might be associated with FRBs. We find one possible association between FRB 180916B and AT2020hur. This paper is organized as follows. In Section \ref{sec:search}, we present the search method and result. In Section \ref{sec:at2020hur}, We give possible explanations for the association between FRB 180916B and AT2020hur. Our discussion and conclusions can be found in Section \ref{sec:dis} and Section \ref{sec:con}.

\section{Search for Astronomical Transients Associated with Fast Radio Bursts} \label{sec:search}

In order to find ATs that may be associated with FRBs, we perform a systematic search for ATs whose positions are consistent with FRBs. The sample of ATs comes from The Open Supernova Catalog (OSC)\footnote{https://sne.space/} and Transient Name Server (TNS)\footnote{https://www.wis-tns.org/}, most of which are supernovae, unclassified optical transients, and a few are gamma-ray bursts.
As of October 1, 2021, there are 90,617 ATs in OSC, 3,771 of which have no coordinate information, and there are 82,798 ATs in TNS. Excluding duplicate sources (55,633), there are a total of 112,915 ATs with certain coordinates.
Recently, CHIME released a catalog of 535 FRBs, which includes 474 nonrepeating FRBs and 62 bursts from 18 previously reported repeaters \citep{CHIME2021}. By October 1, 2021, the total number of fast radio bursts has reached 791, including 587 nonrepeating FRBs and 204 bursts from 22 repeaters\footnote{https://www.wis-tns.org/}.

For each AT-FRB pair, we calculate the distance and the chance possibility between them. Assuming that the spatial distribution of ATs is isotropic, the number of ATs within a specific sky area satisfies the Poisson distribution. The chance probability of finding at least one AT in the vicinity of one FRB is
\begin{equation}
P_{1}=1-\lambda^{0}\exp(-\lambda)/0!=1-\exp(-\lambda),
\end{equation}
where $\lambda=\rho S$ is the expected number of ATs in a given area $S$. The surface number density of ATs is $\rho\approx112915/(41252.96\ \rm{deg^{2}})\approx2.737/\rm{deg^{2}}$. For an AT-FRB pair with distance $D$ (in units of deg), the area can be written as $S\approx[41252.96(1-\cos D)]/2$. To estimate the chance probability conservatively, the distance should include the positional uncertainty of FRB $\delta_{\rm{FRB}}$ and the positional uncertainty of AT $\delta_{\rm{AT}}$. The chance probability of having at least one AT at a distance less than $D$ for all 609 FRBs (including 587 nonrepeating FRBs and 22 repeaters) can be estimated as $P=1-(1-P_{1})^{609}$.

Appendix \ref{sec:pairs} lists the 50 AT-FRB pairs with the most nearest distances. It is found that except for AT2020hur-FRB 180916B pair, all other AT-FRB pairs have a large chance possibility (all close to 100\%), which means that they are unlikely to be associated.
The distance of the second nearest pair is 0.00019 deg. If one neglect the positional uncertainty, the chance possibility would be $P\approx0.02\%$, which corresponds to a $3.7\sigma$ confidence level. However, the derived chance possibility is still close to 100\% when the large positional uncertainty of FRB 20200405A (1.5 deg) is taken into account. The other pairs possess both the large distance (from 0.01 to 0.1 deg) and the large positional uncertainty of FRBs ($\sigma_{\rm{FRB}}\approx0.2$ deg).
Besides, Appendix \ref{sec:pairs} also shows the results of the GRB 110715A-FRB 171209 pair. A detailed discussion on the association between GRB 110715A and FRB 171209 has been presented in \cite{Wang2020b}.

FRB 180916B is a well-localised repeating FRB with $\alpha=01^{\mathrm{h}} 58^{\mathrm{m}} 00.7502^{\mathrm{s}} \pm 2.3$ mas, $\delta=65^{\circ} 43^{\prime} 00.3152^{\prime \prime} \pm 2.3$ mas \citep{Marcote2020}, and AT2020hur is an optical transient with $\alpha=01^{\mathrm{h}} 58^{\mathrm{m}} 00.750^{\mathrm{s}} \pm 1$ arcsec, $\delta=65^{\circ} 43^{\prime} 00.30^{\prime \prime} \pm 1$ arcsec \citep{Lipunov2020}. The distance between FRB 180916B and AT2020hur is 0.0000042 deg (15 mas). Thus, FRB 180916B is well located inside the error circle of AT2020hur.
In order to verify the estimated chance possibility, Monte Carlo simulations are employed. We randomly generate 171,349 ATs and 609 FRBs in the sky. Based on $10^5$ simulations, the probability of having at least one FRB well located inside the error circle of one AT with error radius of 1 arcsec is $\approx0.04\%$, consistent with the analytical estimate, which corresponds to a $3.5\sigma$ confidence level for the AT2020hur-FRB 180916B association.

\begin{figure*}
\centering
\includegraphics[width=0.6\textwidth,angle=0]{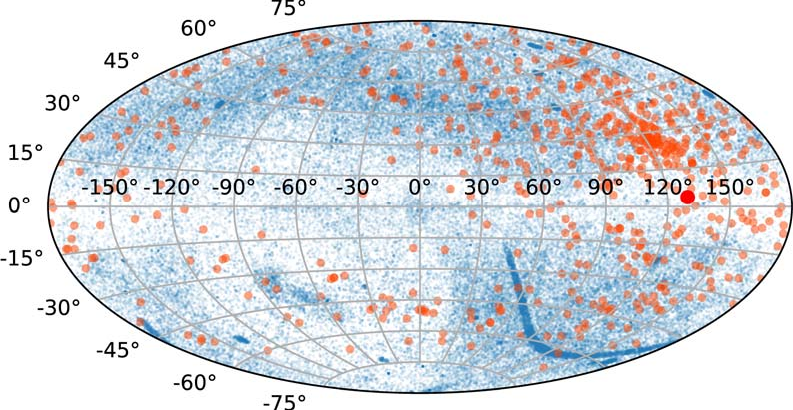}
\caption{The spatial distribution of 112,915 ATs and 609 FRBs that show an anisotropic distribution in the sky. The ATs and FRBs are marked with blue dots and orange circles, respectively. The positions of FRB 180916B and AT2020hur are highlighted with a red circle.
\label{fig:map}}
\end{figure*}

\begin{figure*}
\centering
\includegraphics[width=0.4\textwidth,angle=0]{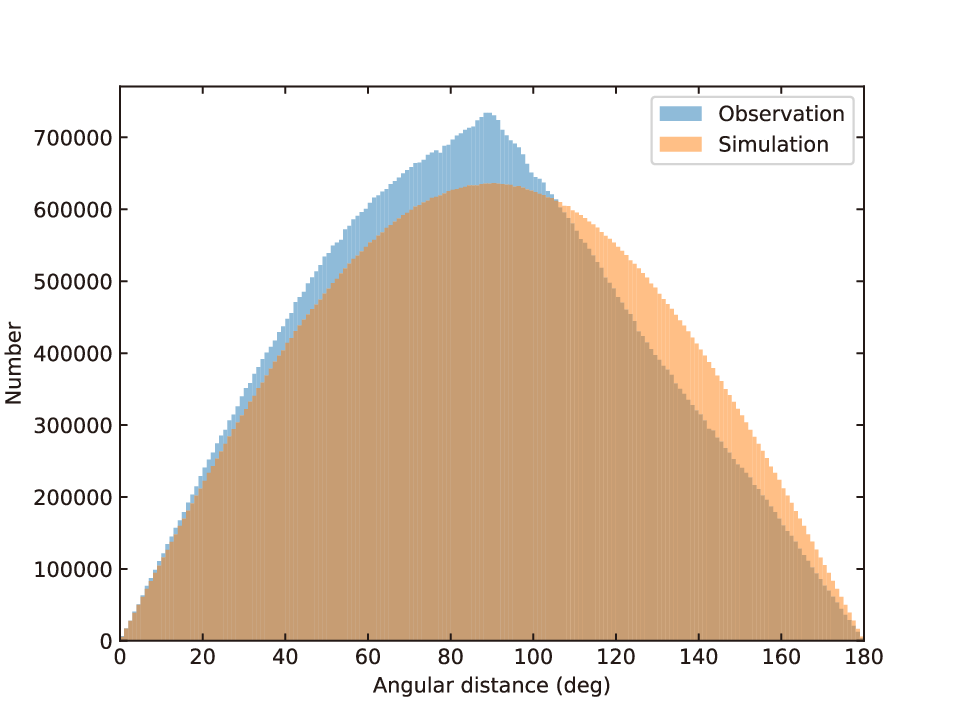}
\includegraphics[width=0.4\textwidth,angle=0]{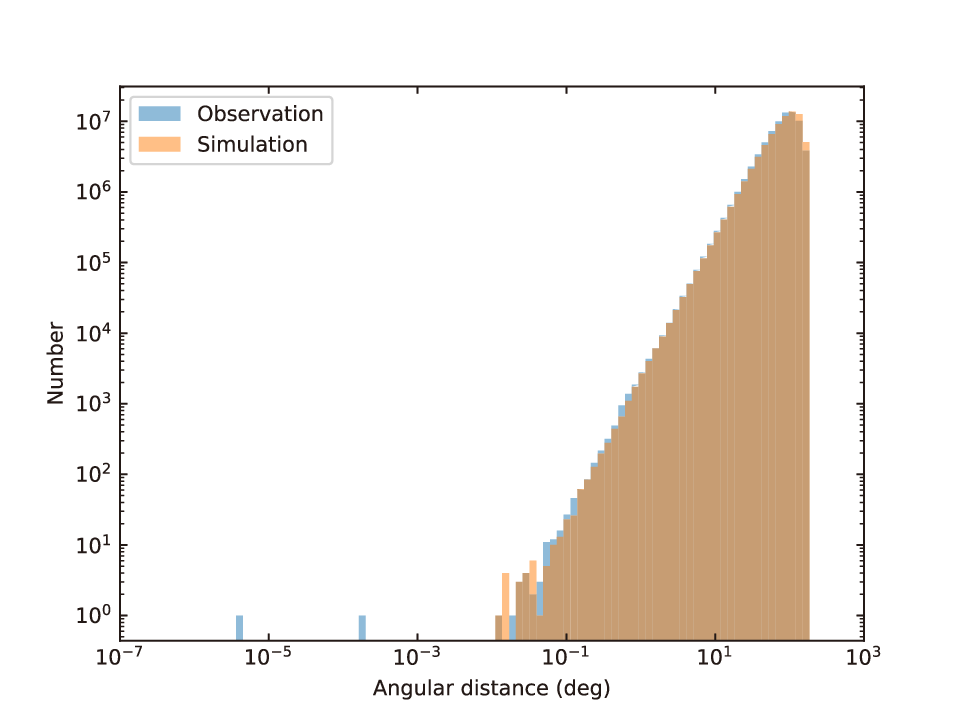}
\includegraphics[width=0.4\textwidth,angle=0]{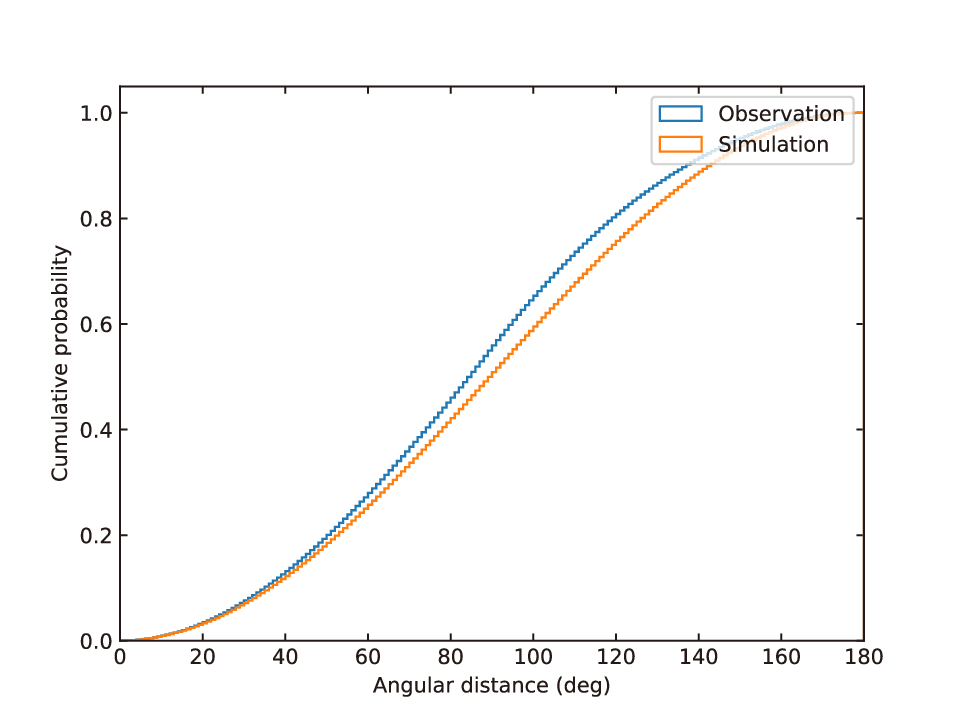}
\includegraphics[width=0.4\textwidth,angle=0]{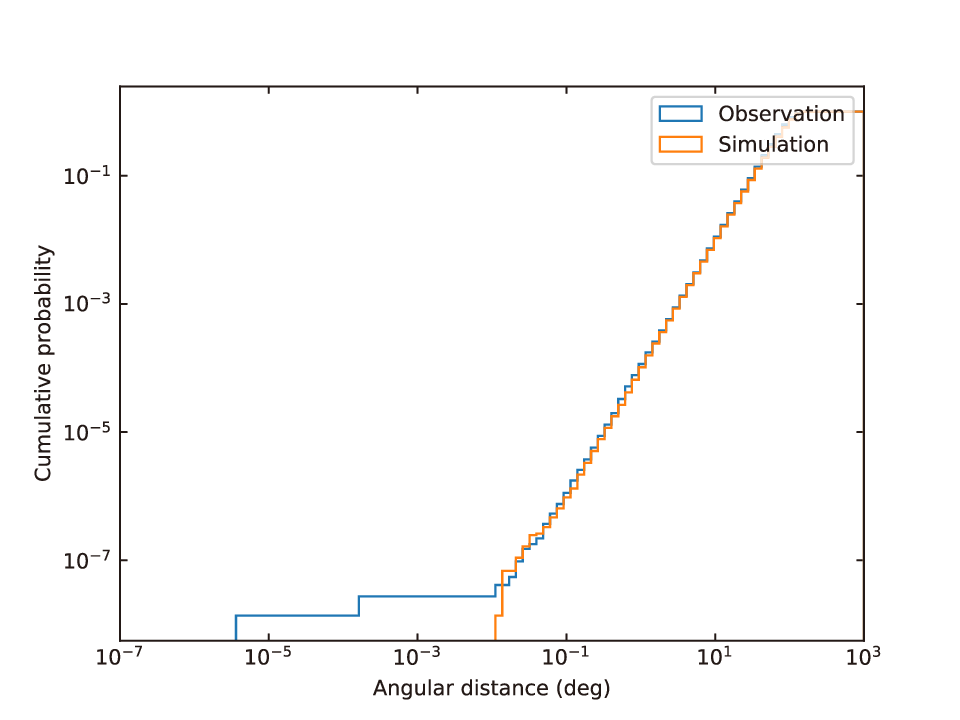}
\caption{The distributions of the angular distance (upper panels) and the cumulative probability distributions of angular distance (lower panels) dervied from the observed data (blue) and the simulated data (orange). The left panels are displayed in linear scale, and the right panels are displayed in logarithmic scale.
\label{fig:cdf}}
\end{figure*}

There are several caveats for the $3.5\sigma$ confidence level.
First, FRB 180916B and AT2020hur has a milliarcsecond and arcsecond precision localization, respectively, and FRB 180916B is well located inside the error circle of AT2020hur. One may directly use the angular distance between FRB 180916B and AT2020hur to calculate the chance possibility. The derived chance possibility is $P=0.00001\%$ which corresponds to a $5.34\sigma$ confidence level.
Second, the spatial distribution of the 112,915 ATs is not isotropic, as shown in Figure \ref{fig:map}. In Figure \ref{fig:cdf}, we plot the distribution of angular distance and cumulative probability distribution of angular distance dervied from the observed data and simulated data. The deviation of the observed data from the simulated data may be due to the anisotropic distribution of FRBs and ATs. We can use the effective density of ATs near FRBs to calculate the chance probability, i.e. $P=1-\prod_{i=1}^{609}(1-P_{1,i})$, where $P_{1,i}=1-\exp(-\rho_{i} S)$. For each FRB, we count the number of ATs $N_{i}$ whose angular distance from it is less than 10 deg (which corresponds to a solid angle of about 313 deg$^{2}$), thus the effective surface density of ATs near each FRB is $\rho_{i}\approx N_{i}/313$. We update the chance probability to 0.0438\% ($3.5\sigma$), or 0.00001\% ($5.34\sigma$) if one directly use the angular distance between FRB 180916B and AT2020hur to calculate the chance possibility, which is is consistent with the results derived by assuming an isotropic distribution.
Third, as shown in Figure \ref{fig:multi}, AT2020hur occurs in one of the active windows of FRB 180916B, which is an independent support for the association.

\section{AT2020hur: A Possible Optical Counterpart of FRB 180916B} \label{sec:at2020hur}

\subsection{Observational Properties of AT2020hur}

AT2020hur is an optical transient which was first discovered on 2020-04-08 at 23:18:41.184 (MJD=58947.97130787) by MASTER-Kislovodsk robotic telescope \citep{Lipunov2020}. The unfiltered magnitude at the time of discovery was 18.4 mag (Vega system; \citealt{Lipunov2020}). On 2020-04-09 at 19:07:30 (MJD=58948.79687500), about 0.8 days after its discovery, MASTER-Kislovodsk again reported an optical observation of AT2020hur, which was as bright as when it was discovered \citep{Lipunov2020}.

We follow the procedure of \cite{Gorbovskoy2012} to convert unfiltered magnitudes of MASTER telescope into fluxes. The calculated Vega flux in the CCD spectral band of MASTER telescope is $F_{o}^{W}=\int F_{\rm{Vega}}(\lambda)W(\lambda)d\lambda=1.33\times10^{-5}\ \rm{erg\ cm^{-2}\ s^{-1}}$. The unfiltered magnitude can be converted into the absolute flux value by using the Pogson equation $F^{W}=F_{o}^{W}\times10^{-0.4 W}$. The flux density at the wavelength of the CCD response maximum ($5500\, \rm{\AA}$) can derived by dividing the flux by an effective frequency interval $\Delta\nu_{\rm{eff}}\approx3.9\times10^{14}\ \rm{Hz}$ of the CCD response function. With above equations, one can derive that the optical flux for MASTER unfiltered magnitude 18.4 mag is about $5.8\times10^{-13}\ \rm{erg\ cm^{-2}\ s^{-1}}$, and the flux density at $5500\ \rm{\AA}$ is about 0.15 mJy.

If AT2020hur is indeed associated with FRB 180916B, according to the redshift $z=0.0337$ of the host galaxy of FRB 180916B \citep{Marcote2020}, which corresponds to a luminosity distance of $D_L=153.7$ Mpc using the cosmological parameters from Planck 2018 results \citep{Planck2020}, the isotropic luminosity of AT2020hur is about $1.64\times10^{42}\ \rm{erg\ s^{-1}}$. For the first and second optical observations, the exposure times are 290 s and 60 s, thus the average isotropic energy released during the two exposures are $4.60\times10^{44}\ \rm{erg}$ and $9.52\times10^{43}\ \rm{erg}$, respectively. Assuming a constant luminosity over the 0.8 days, the isotropic energy released during this interval is about $1.13\times10^{47}\ \rm{erg}$.

\subsection{Fitting AT2020hur with Giant Flare Afterglow Model}

One possible explanation for the AT2020hur-FRB 180916B association, is that FRB 180916B is powered by a flaring magnetar, as suggested by the popular FRB models, while AT2020hur originates from the afterglow of one or more energetic giant flares (GFs). Relativistic outflow can be launched during the GF. When the relativistic outflow propagates outward and interacts with the surrounding medium, a pair of shocks is formed. The forward shock (FS) propagates into the surrounding medium, and the reverse shock (RS) propagates into the outflow. If the magnetization parameter of the outflow is high enough, The short-lived RS can be suppressed \citep{Zhang2005,Mimica2009,Mizuno2009}. Here we only consider the contribution of emission from the FS.

\begin{figure*}
\centering
\includegraphics[width=0.4\textwidth,angle=0]{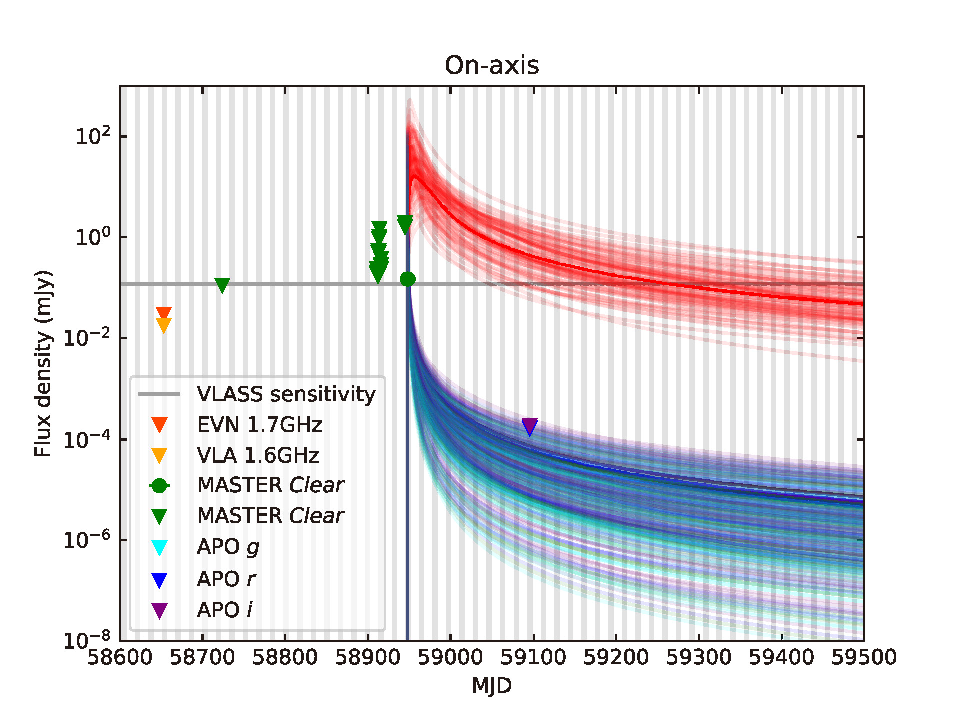}
\includegraphics[width=0.4\textwidth,angle=0]{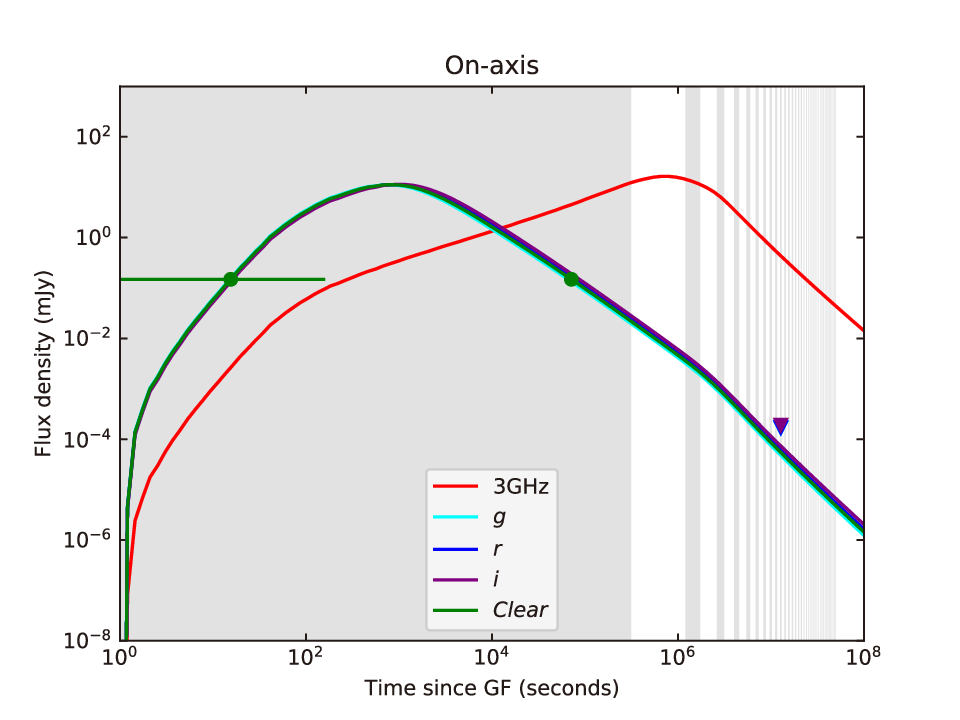}
\includegraphics[width=0.4\textwidth,angle=0]{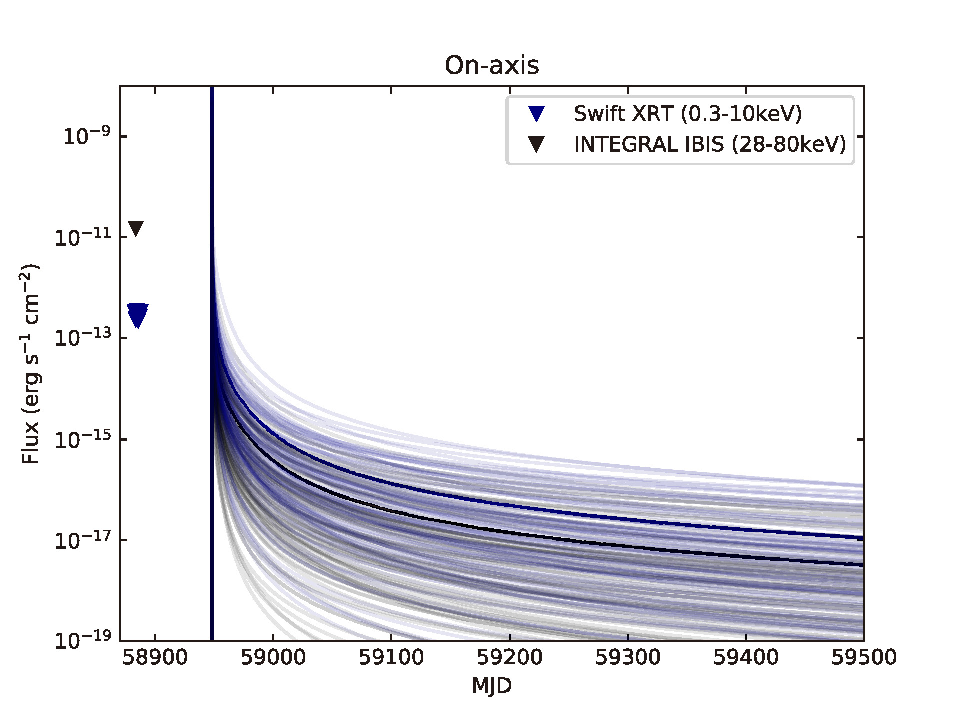}
\includegraphics[width=0.4\textwidth,angle=0]{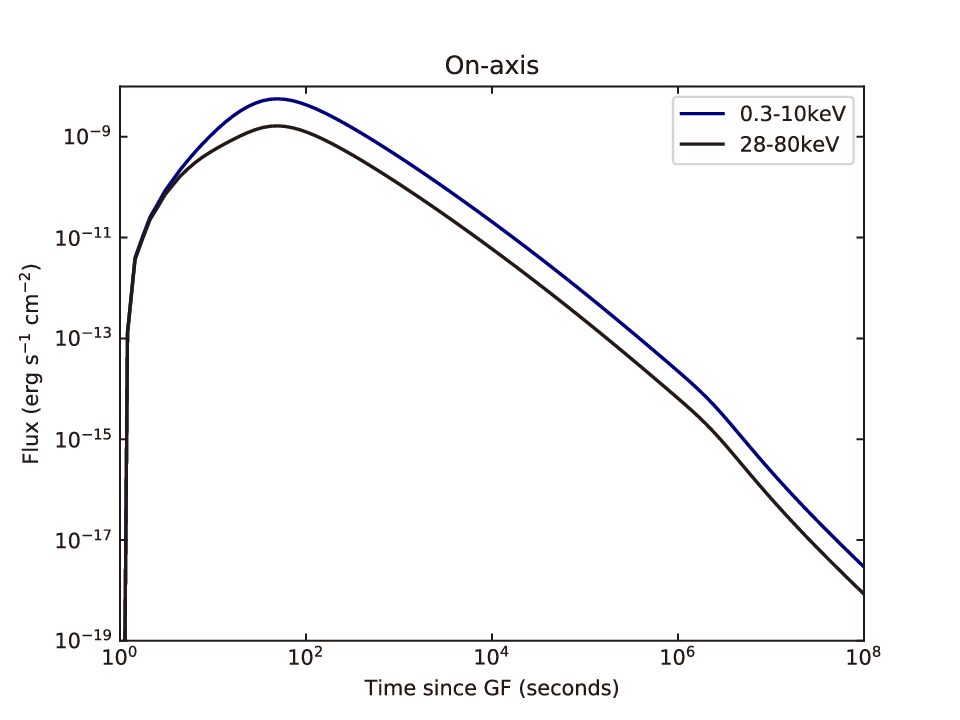}
\caption{The standard forward shock afterglow model fit to AT2020hur and multiwavelength observations of FRB 180916B. Upper left: the optical and radio lightcurves in a linear time scale. Green circles are the $C$-band observations of AT2020hur \citep{Lipunov2020}. Green upside down triangles are the $C$-band upper limits of FRB 180916B \citep{Zhirkov2020}. Cyan, blue and purple upside down triangles are the $g$-, $r$-, and $i$-band upper limits of FRB 180916B observed by Apache Point Observatory (APO; \citealt{Kilpatrick2021}). Orange and yellow upside down triangles are the upper limits on the persistent radio emission associated with FRB 180916B observed by the European Very-long-baseline-interferometry Network (EVN) and the Karl G. Jansky Very Large Array (VLA; \citealt{Marcote2020}). Gray horizontal line is the single epoch sensitivity of the 3 GHz VLA Sky Survey (VLASS; \citealt{Lacy2020}). Upper right: the optical and radio lightcurves in a logarithmic times cale. Lower left: the X-ray and $\gamma$-ray lightcurves in a linear time scale. Navy upside down triangles are the X-ray upper limits observed by Swift XRT \citep{Tavani2020a}. Black upside down triangle is the $\gamma$-ray upper limits observed by INTEGRAL IBIS \citep{Panessa2020}. Lower right: X-ray and $\gamma$-ray lightcurves in a logarithmic time scale. The gray shaded regions in the upper left and right panels correspond to the predicted activity days of FRB 180916B for a period of 16.29 days and a 6.1 day activity windows \citep{Pastor2020}
\label{fig:multi}}
\end{figure*}

We develop a standard forward shock afterglow numerical model, in which the dynamical evolution of the outflow follows \cite{Huang1999,Huang2000}, and synchrotron radiation from the electrons with a segmented power-law distribution \citep{Sari1998} is invoked to calculate the afterglow lightcurves. Both on-axis and off-axis configurations are considered in our model. The free parameters in our model include:
the half-opening angle of the outflow $\theta_j$ (since the outflow can be anisotropic),
the viewing angle $\theta_v$ (valid only in off-axis configuration, and $\theta_v>\theta_j$ is required),
the isotropic kinetic energy of the outflow $E_{\rm{K,iso}}$,
the initial Lorentz factor of the outflow $\Gamma_0$,
the number density of the surrounding medium $n$ (considering a constant medium density),
the fraction of the shock internal energy that is partitioned to magnetic fields $\epsilon_B$,
the fraction of the shock internal energy that is partitioned to electrons $\epsilon_e$,
the electron energy spectral index $p$,
and the time interval between the discovery time of AT2020hur and the launch time of the assumed GF $t_{\rm{shift}}$.

The Markov Chain Monte Carlo (MCMC) method implemented in the emcee Python package \citep{emcee} is employed to determine the posterior probability distributions and the best-fit parameter values.
We set a wide enough range for the priors in order to explore a parameter space as large as possible, except for $E_{\rm{K,iso}}$.
To date, there are four GFs and two GF candidates have been discussed in the literature. The most energetic GF releases a total isotropic energy $E_{\rm{\gamma,iso}}\sim10^{47}\ \rm{erg}$ \citep{Frederiks2007,Yang2020b,ZhangH2020}. The isotropic kinetic energy of the outflow can be written as $E_{\rm{K,iso}}=E_{\rm{\gamma,iso}}(1/\eta-1)$. The radiative efficiency $\eta$ depending on the specific radiation mechanism has a large uncertainty. Observationally, the GRB radiative efficiency is found to vary from less than 0.1\% to over 90\% \citep{Fan2006,Zhang2007,Wang2015}. If GF and GRB share the similar radiation mechanisms, GF may have the same distribution of $\eta$. Conservatively, we set the prior of $E_{\rm{K,iso}}$ to range from $10^{47}\ \rm{erg}$ to $10^{50}\ \rm{erg}$.

In addition to the two optical points reported by the MASTER, we also searched for other multiwavelength observations of FRB 180916B. The multiwavelength data used to fit are shown in Figure \ref{fig:multi}. Figures \ref{fig:multi} also shows the fitting results from on-axis configuration. In Appendix \ref{sec:multi2}, we show the fitting results from off-axis configuration. We find that the overall quality of the fitting is good for both configurations, indicating that the standard forward shock afterglow model can interpret the lightcurve of AT2020hur. Table \ref{tab:par} shows the parameters, priors, and fitting results of our model. Appendix \ref{sec:corner} shows the one and two dimensional projections of the posterior probability distributions of parameters with the corner plots. As shown in the corner plots, the posterior probability distributions of several parameters is dispersed due to the lack of the data. However, the posterior probability distributions of $E_{\rm{K,iso}}$ are gathered and close to the upper limit of its prior, i.e., $10^{50}\ \rm{erg}$, which suggests a large kinetic energy of the outflow. For the on-axis configuration, the isotropic kinetic energy of the outflow ranges from $2.3\times10^{49}$ to $7.9\times10^{49}\ \rm{erg}$, and the half-opening angle of the outflow ranges from 0.68 to 2.59 rad, thus the true (beaming-corrected) kinetic energy of the outflow $E_{\rm{K}}=E_{\rm{K,iso}}(1-\cos\theta_j)/2$ ranges from $2.5\times10^{48}$ to $7.4\times10^{49}\ \rm{erg}$. For the off-axis configuration, $E_{\rm{K}}$ ranges from $8.0\times10^{48}$ to $7.9\times10^{49}\ \rm{erg}$. According to the derived initial Lorentz factor ranging from 14 (15) to 71 (91), the total mass of the outflow $M_{\rm{ej}}=E_{\rm{K}}/[(\Gamma_0-1)c^2]$ ranges from $4\times10^{25}$ ($1\times10^{26}$) to $6\times10^{27}\ \rm{g}$ ($6\times10^{27}\ \rm{g}$) for the on-axis (off-axis) configuration.

\begin{deluxetable}{lcrr}\label{tab:par}
\setlength\tabcolsep{1pt}
\tablecaption{Free parameters, priors, and best-fit results in our model.}
\tablehead{
\colhead{Parameter}	&	\colhead{Prior}	&	\colhead{Result (on-axis)}	&	\colhead{Result (off-axis)}
}
\startdata
$\theta_v\rm\ (rad)$	&	$[0,\pi]$	&	-	&	$1.97_{-0.91}^{+0.81}$	\\
$\theta_j\rm\ (rad)$	&	$[0,\pi]$	&	$1.56_{-0.88}^{+1.03}$	&	$1.83_{-0.91}^{+0.77}$	\\
$\log[E_{\rm K,iso}\rm\ (erg)]$	&	$[47,50]$	&	$49.68_{-0.32}^{+0.22}$	&	$49.80_{-0.19}^{+0.13}$	\\
$\log\Gamma_0$	&	$[1,3]$	&	$1.44_{-0.29}^{+0.41}$	&	$1.48_{-0.31}^{+0.48}$	\\
$\log[n\rm\ (cm^{-3})]$	&	$[-6,3]$	&	$1.44_{-1.29}^{+1.01}$	&	$1.95_{-0.80}^{+0.60}$	\\
$\log\epsilon_{B}$	&	$[-7,-0.5]$	&	$-1.15_{-0.71}^{+0.48}$	&	$-1.10_{-0.53}^{+0.39}$	\\
$\log\epsilon_{e}$	&	$[-7,-0.5]$	&	$-0.79_{-0.35}^{+0.21}$	&	$-0.72_{-0.19}^{+0.14}$	\\
$p$	&	$[2,3]$	&	$2.54_{-0.25}^{+0.28}$	&	$2.53_{-0.18}^{+0.22}$	\\
$\log[t_{\rm{shift}}\rm\ (s)]$	&	$[0,7]$	&	$1.11_{-0.62}^{+0.83}$	&	$3.38_{-0.72}^{+0.70}$	\\
\enddata
\tablecomments{The uncertainties of the best-fit parameters are measured as $1\sigma$ confidence ranges.}
\end{deluxetable}

\section{Discussion} \label{sec:dis}
\subsection{Can Giant Flare Afterglow Model Explain AT2020hur?}

Although the giant flare afterglow model can interpret the observations of AT2020hur and multiwavelength constraints on FRB 180916B, there are some issues for this model.

First, the derived isotropic kinetic energy is $\sim 10^{49}-10^{50}\ \rm{erg}$, which is more than 3 orders of magnitude larger than the kinetic energy inferred by the typical GFs. For example, GRB 200415A is one of the most energetic extragalactic GFs with an isotropic energy released in the initial pulse and following tail of $\sim10^{46}$ erg \citep{Yang2020b,ZhangH2020,CastroTirado2021,FermiLAT2021,Roberts2021,Svinkin2021}. GRB 200415A is also the first GF accompanied by GeV emission which is generally believed to originate from the GF afterglow \citep{Yang2020b,ZhangH2020,FermiLAT2021}. The inferred kinetic energy of the outflow is comparable to the total radiated energy of GRB 200415A \citep{ZhangH2020,FermiLAT2021}. Besides, it is suggested that the outflow is ultra-relativistic with a bulk Lorentz factor of $\sim100$ \citep{ZhangH2020,FermiLAT2021}, which is comparable to our results. However, our derived baryon loading is three orders of magnitude larger than the estimated baryon loading in the relativistic outflow of GRB 200415A, i.e. $\sim10^{23}$ g \citep{ZhangH2020}. Another two galactic GFs from soft gamma-ray repeaters SGR 1806-20 and SGR 1900+14 were accompanied by radio afterglows \citep{Frail1999,Cameron2005,Gaensler2005}, and the inferred kinetic energy are of the order of $10^{44}-10^{46}$ erg (e.g. \citealt{Cheng2003,Dai2005,Ioka2005,Nakar2005,WangXY2005}). For GF from SGR 1900+14, The extremely high luminosity, hard spectrum and short duration suggest that a relativistic fireball ($\Gamma\gtrsim10$) with very low baryon contamination is lauched by a neutron star with magnetic field of $10^{15}$ G \citep{Thompson2001}. Obviously, if AT2020hur is indeed powered by an energetic GF afterglow, the GF doesn't belong to the class of typical GFs. If the large kinetic energy is powered by the internal magnetic energy dissipation, one can place an lower limit on internal magnetic field $B\gtrsim(6E_{\rm K}/R_{\rm NS}^{3})^{1/2}\sim8\times10^{15}E_{\rm K,49}^{1/2}R_{\rm NS,6}^{-3/2}$ G, where $R_{\rm NS}$ is the neutron star radius, $E_{\rm K,49}=E_{\rm K}/10^{49}$ erg, $R_{\rm NS,6}=R_{\rm NS}/10^{6}$ cm.

\begin{figure}
\centering
\includegraphics[width=0.4\textwidth,angle=0]{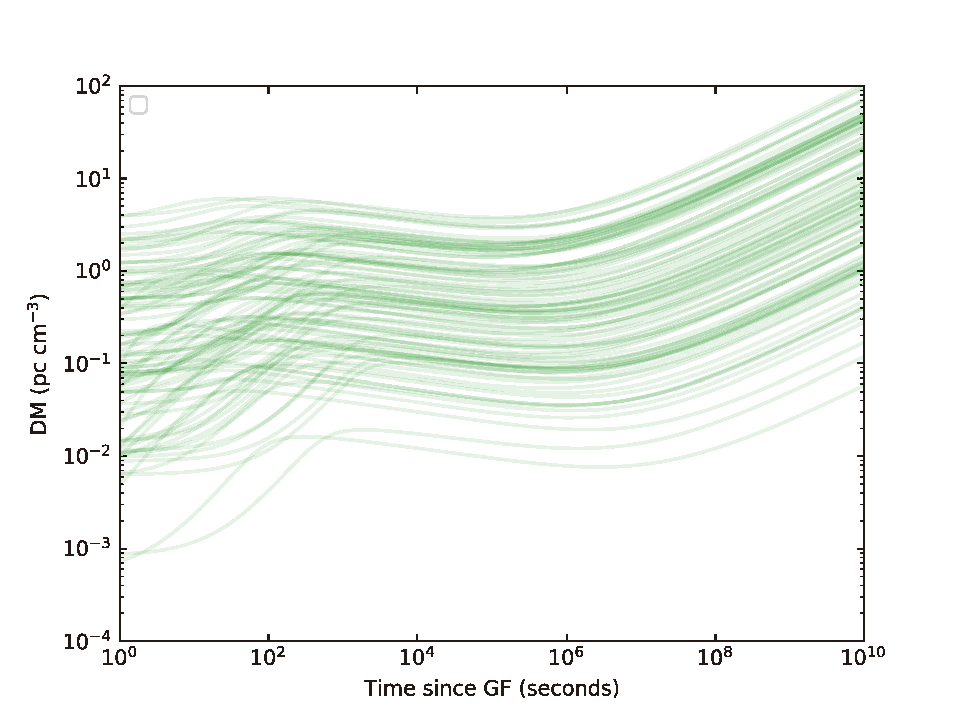}
\includegraphics[width=0.4\textwidth,angle=0]{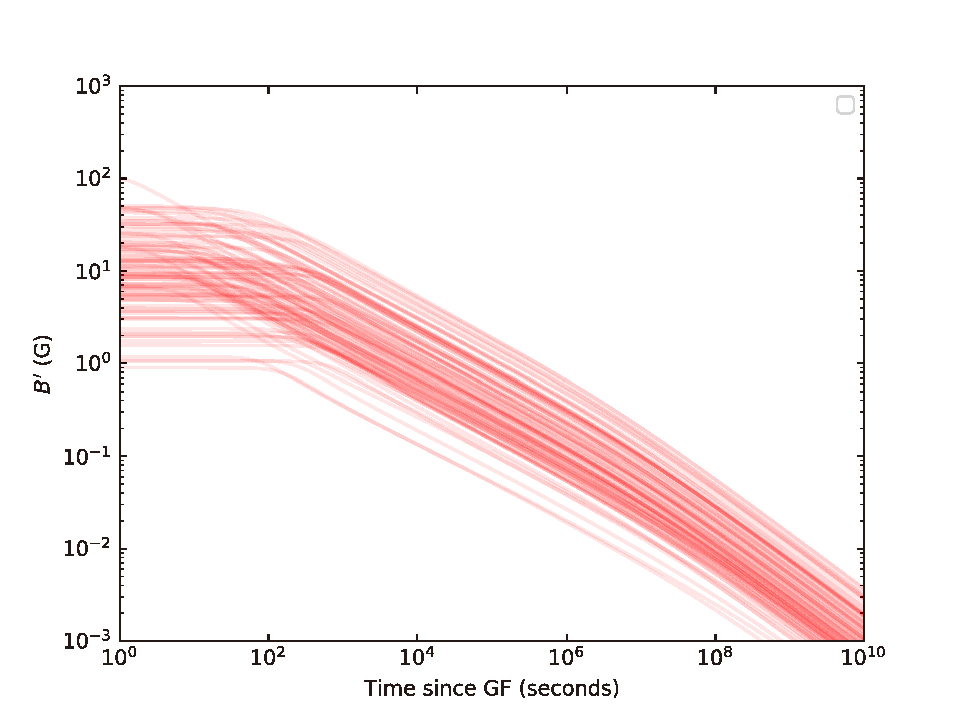}
\includegraphics[width=0.4\textwidth,angle=0]{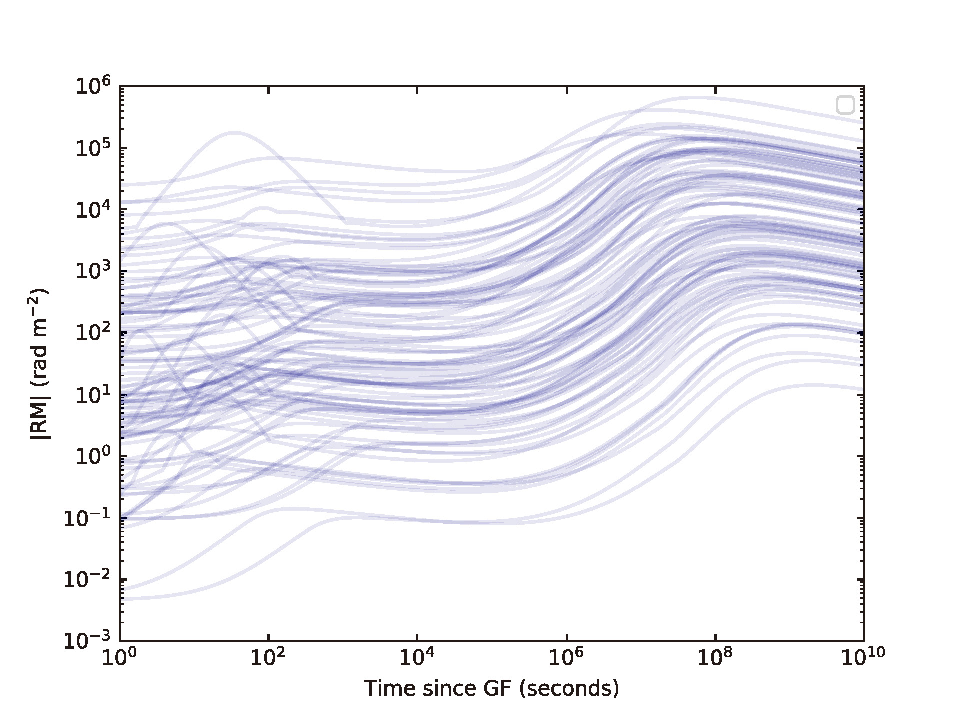}
\caption{Time evolution of the DM, magnetic field, and RM from the shocked material, derived from a sample of parameters within the $1\sigma$ posterior probability distributions.
\label{fig:dm}}
\end{figure}

Second, for the on-axis configuration, the contribution of the shocked material to the DM and rotation measure (RM) of FRB 180916B must be taken into account. There are four regions when the relativistic outflow interacts with the sourrounding medium: (1) the unshocked surrounding medium, (2) the shocked surrounding medium, (3) the shocked outflow, and (4) the unshocked outflow. The two shocked, ionized regions provide free electrons and magnetic field to affect the DM and RM. For the long-term evolution of the DM and RM, the contribution from the the shocked outflow can be neglected, since the shocked material is dominated by the shocked surrounding medium. The DM of the shocked surrounding medium is given by \citep{Yu2014}
\begin{equation}\label{DM}
\mathrm{DM_s}=\int \frac{\mathcal{D}}{1+z} n_{s}^{\prime} d l_{s}^{\prime}=\frac{\mathcal{D}}{1+z}n_{s}^{\prime} \Delta R_{s}^{\prime},
\end{equation}
where $\mathcal{D} \equiv [\Gamma(1-\beta \cos \theta)]^{-1}$ is the the Doppler factor due to the relativistic motion of the shocked material, $n_{s}^{\prime}$ is the comoving number density of the shocked material, and $\Delta R_{s}^{\prime}$ is the comoving width of the shocked material (in this paper, we use the superscript prime to denote the quantities in the shock comoving frame). Here we have neglected the internal structure of the blast wave \citep{Blandford1976}. According to the jump condition $n_{s}^{\prime}/n=(\hat{\gamma} \Gamma+1)/(\hat{\gamma}-1)$ (where $\hat{\gamma}$ is the adiabatic index, and $\Gamma$ is the Lorentz factor of the blast wave), $\Delta R_{s}^{\prime}$ can be written as
\begin{equation}
\Delta R_{s}^{\prime} = \frac{R(\hat{\gamma}-1)}{3(\hat{\gamma} \Gamma+1)},
\end{equation}
where $R$ is the blast wave radius.
The RM for a plasma shocked by relativistic shock can be written in the form
\begin{equation}
\begin{aligned}
\mathrm{RM_s}=&\frac{e^{3}}{2 \pi m_{e}^{2} c^{4}} \frac{\mathcal{D}^{2}}{(1+z)^{2}} B_{\|}^{\prime} \Delta R_{s}^{\prime}\times\\
&\int_{\min(\gamma_{m}^{\prime},\gamma_{c}^{\prime})}^{\gamma_{M}^{\prime}} \max \left(\frac{1}{\gamma_{e}^{\prime 2}}, \frac{\ln \gamma_{e}^{\prime}}{2 \gamma_{e}^{\prime 2}}\right) \frac{d N_{e}^{\prime}}{d \gamma_{e}^{\prime}} d \gamma_{e}^{\prime}\\
=&0.812 \frac{\mathcal{D}^{2}}{(1+z)^{2}} \left(\frac{B_{\|}^{\prime}}{\mathrm{\mu G}}\right) \left(\frac{\Delta R_{s}^{\prime}}{\mathrm{pc}}\right)\times\\
&\int_{\min(\gamma_{m}^{\prime},\gamma_{c}^{\prime})}^{\gamma_{M}^{\prime}} \max \left(\frac{1}{\gamma_{e}^{\prime 2}}, \frac{\ln \gamma_{e}^{\prime}}{2 \gamma_{e}^{\prime 2}}\right) \frac{d N_{e}^{\prime}}{d \gamma_{e}^{\prime}} d \gamma_{e}^{\prime}\ \mathrm{rad\ m^{-2}},
\end{aligned}
\end{equation}
where $B_{\|}^{\prime}$ is the component of the magnetic field along the line-of-sight, $d N_{e}^{\prime}/d \gamma_{e}^{\prime}$ is the electron distribution of the shocked material, and $\gamma_{m}^{\prime}$, $\gamma_{c}^{\prime}$, and $\gamma_{M}^{\prime}$ are the corresponding minimum Lorentz factor, the cooling Lorentz factor, the maximum Lorentz factor. The $\ln \gamma_{e}^{\prime}/(2 \gamma_{e}^{\prime 2})$ term accounts for the suppression of the RM contributed by ultra-relativistic electrons \citep{Quataert2000}, and the $1/\gamma_{e}^{\prime 2}$ term is an approximate correction factor for the RM contributed by electrons between non-relativistic and ultra-relativistic.

Figure \ref{fig:dm} shows the evolution of $\mathrm{DM_s}$, $B^{\prime}$, and $\mathrm{RM_s}$, derived from a sample of parameters within the $1\sigma$ posterior probability distributions. Due to the poor constraints on the parameters, the distribution of DM spans a wide range, from $\sim10^{-3}$ to $\sim10^{2}\ \mathrm{pc\ cm^{-3}}$. Observationally, the DM of FRB 180916B is about $350\ \mathrm{pc\ cm^{-3}}$. Excluding the contribution of our Galaxy, the excess DM in this line of sight is estimated to be about $150\ \mathrm{pc\ cm^{-3}}$ or $25\ \mathrm{pc\ cm^{-3}}$, base on the NE2001 or YMW2016 model \citep{Cordes2002,Cordes2003,Yao2017}. The excess DM should be treated as a stringent upper limit of the $\mathrm{DM_s}$. As shown in Figure \ref{fig:dm}, $\mathrm{DM_s}\lesssim25\ \mathrm{pc\ cm^{-3}}$ alway holds within $\lesssim10^9$ seconds ($\lesssim30$ years) since the GF. Since $\mathrm{DM_s}$ increases slowly at late time, future observations of such a trend may help verify our model\footnote{Previous observations of FRB 180916B did not indicate a significant variation in DM (e.g. \citealt{CHIME2020a,Nimmo2021,Sand2021}), which is consistent with our model.}. The distribution of $\mathrm{RM_s}$ also spans a wide range, from $\sim10^{-2}$ to $\sim10^{6}\ \mathrm{rad\ m^{-2}}$. \cite{CHIME2019} reported that FRB 180916B has a measured RM of $-114.6\pm0.6\ \mathrm{rad\ m^{-2}}$. The $\mathrm{RM_{MW}}$ contributed by the Milky Way along this line of sight is $\mathrm{RM_{MW}}\approx-72\pm23\ \mathrm{rad\ m^{-2}}$ \citep{Oppermann2015} or $\mathrm{RM_{MW}}\approx-115\pm12\ \mathrm{rad\ m^{-2}}$ \citep{Ordog2019}, which suggests that the excess RM only ranges from zero to several tens of $\mathrm{rad\ m^{-2}}$. As shown in Figure \ref{fig:dm}, due to a strong magnetic field amplified by the relativistic shock, the derived $\mathrm{RM_s}$ usually too large with respect to the excessed DM, especially at late time when the relativistic electrons are so few that $\mathrm{RM_s}$ cannot be suppressed efficiently. However, there are still a small fraction of $\mathrm{RM_s}$ satisfy the observational requirements, which means the on-axis configuration is still possible. For the off-axis configuration, since there is no shocked material contributing to the DM and RM along the line of sight, the standard forward shock afterglow with the off-axis geometry always holds.

Third, the single GF afterglow model suggests that the first optical detection corresponds to the first tens or hundreds of seconds of the rising phase of the GF afterglow. This is unlikely since the rise time of the afterglow is only a small part of the total afterglow duration. For example, if an optical telescope detects such a GF afterglow, the probability of detecting the optical afterglow within $\sim10-100$ seconds after the GF is $\sim0.01-0.1\%$ (assuming that the duration of the afterglow above the fluence threshold is $\sim10^{5}$ seconds).

Within the framework of the giant flare afterglow model, AT2020hur might be associated with multiple giant flares. Assuming that the two observation times of AT2020hur correspond to the peak times of two giant flare afterglows, one can find \citep{Sari1998}
\begin{equation} \label{eq:flux}
F_{\nu, \max }=1.1 \times 10^{5} \epsilon_{B}^{1 / 2} E_{\mathrm{K,iso,52}} n^{1/2} D_{L,28}^{-2}\ \mu \mathrm{Jy}\approx150\ \mu \mathrm{Jy},
\end{equation}
where $E_{\mathrm{K,iso,52}}=E_{\mathrm{K,iso}}/10^{52}\ \mathrm{erg}$, $D_{L,28}=D_{L}/10^{28}\ \mathrm{cm}$. For AT2020hur,  the derived isotropic kinetic energy of the outflow $E_{\mathrm{K,iso}}\approx3.1\times 10^{46} \epsilon_{B}^{-1 / 2} n^{-1 / 2}\ \mathrm{erg}$, which is about $2-3$ orders of magnitude lower than the one derived from the single giant flare afterglow model for typical parameters (e.g. $\epsilon_{B}=0.1$, $n=1\ \mathrm{cm^{-3}}$). Although the multiple GF afterglow model derives a much lower kinetic energy, it would be more difficult for a magnetar to produce multiple energetic GFs in less than a day. Besides, if there are two GFs accounting for the optical counterpart, two optical detections correspond to the peaks of the GF optical afterglow, which would have an extremely lower probability for such a picture. Unless the multiple GF afterglow model has an off-axis configuration, the contribution of the shocked material induced by each GF to the DM and RM of FRB 180916B may increase significantly {\em with the increase of GF}. In order to be consistent with the small observed DM and RM for FRB 180916B, smaller values of  $n$ and $\epsilon_{B}$ are needed. However, this results in a larger kinetic energy (see Equation \ref{eq:flux}).

We therefore conclude that the giant flare afterglow model is unlikely to explain the optical counterpart of FRB 180916B, AT2020hur, since there is a lot of fine tuning and coincidences required for this specific model. As shown in Figure \ref{fig:multi} and \ref{fig:multi2}, the radio flux density is still likely to be above the single-epoch sensitivity of 3 GHz VLA Sky Survey (VLASS; \citealt{Lacy2020}) in the future. Thus, further observations of the radio counterpart of FRB 180916B may help verify the giant flare afterglow model.

\subsection{Other possible origins of AT2020hur}

We here discuss the other possible origins of AT2020hur. Due to FRB 180916B is a repeating FRB, the optical counterparts associated with the catastrophic FRB models are unlikely to be the origin of AT2020hur. For instance, a type Ia supernova associated with an FRB produced by the double white dwarf merger \citep{Kashiyama2013}, a GRB afterglow or a kilonova associated with an FRB produced by the binary neutron star mergers (e.g. \citealt{Totani2013,WangJS2016,WangJS2018}), or a GRB associated with an FRB produced by a neutron star collapses to a black hole \citep{ZhangB2014}. AT2020hur is also unlikely to originate from the progenitor systems of FRBs, like a supernova may have a very long time delay with respect to the FRB itself which is inconsistent with the case of AT2020hur and FRB 180916B.

For the optical counterparts arising from the circumburst environments, a ``cosmic comb" model is a promising model that can produce both FRBs and optical counterparts, the optical counterparts can originate from a variety of ``combs", including AGN, GRB, SN, etc \citep{ZhangB2017}. However, this model cannot explain the periodicity of FRB 180916B. The binary comb model is an upgraded version of the ``cosmic comb" model that can explain periodic FRBs, but the model does not predict an optical counterpart \citep{Ioka2020}. Although \cite{Yang2021} suggested that an optical transient is expected from FRBs heating companion star in close binary system, the optical emission is so weak that only the counterparts of those galactic FRBs can be detected \citep{Yang2021}.

We next consider the prompt FRB counterparts which usually have short durations ($\lesssim$ 100 seconds), as discussed and summarized in \cite{Chen2020}. For \cite{Metzger2019} model, a weak optical counterpart is expected if the upstream plasma is composed of electrons and positrons, however, there is no quantitative prediction about the luminosity and duration of the optical counterpart. For \cite{Beloborodov2020} model, a bright optical flash is produced when the blast wave strikes the hot wind bubble, and it is expected that the duration of the optical flash is $\lesssim$ 1 seconds and the upper limit of energy released is $\sim10^{44}$ ergs, which is comparable to the isotropic energy released during the two exposures of AT2020hur.

Another possible origin of AT2020hur is that it may be two fast optical bursts (FOBs) produced by the inverse Compton (IC) scattering of FRB emissions, as suggested by \cite{Yang2019b}. The IC scattering process can occur in the FRB emission region, i.e. in the pulsar magnetosphere or in the maser outflow. For the ``one-zone" IC scattering model, the optical flux of the FOB is $\sim 5 \times(10^{-13}-10^{-2})\ \mathrm{Jy}$ or $\lesssim 1.6 \times 10^{-6}\ \mathrm{Jy}$ for FOB and FRB both formed in the pulsar magnetosphere or in the maser outflow, and the duration of the FOB is $\sim1$ ms in these two cases. One can find that the energy released in the optical band is $\sim10^{30}-10^{41}\ \mathrm{erg}$ or $\lesssim 5 \times 10^{36}\ \mathrm{erg}$, respectively, which don't meet the energy requirement of AT2020hur. The IC scattering process can also occurs in a different region from the FRB emission, e.g. IC scattering region is in the nebula, while the FRB is formed via coherent radiation near the neutron star. For the ``two-zone" IC scattering model, one has the upper limit of the FOB flux $\lesssim 8.8 \times 10^{-3}\ \mathrm{Jy}$ and the duration of the FOB $\sim 5 \times 10^{3}$ s, which corresponds to the energy released is $\lesssim 10^{47}\ \mathrm{erg}$. Therefore the ``two-zone" IC scattering model is a promising model to explain the AT2020hur. \cite{Yang2019b} also discussed the FOB produced by the same emission mechanism as FRB. however, the derived flux in the optical band is extremely low.

\subsection{The coincidence between AT2020hur and activity window of FRB 180916B}

In Figure \ref{fig:multi}, we plot the lightcurve of AT2020hur and predicted activity days of FRB 180916B for a period of 16.29 days and a 6.1 day activity windows \citep{Pastor2020}. As shown in Figure \ref{fig:multi}, AT2020hur is located in one of the activity windows (MJD from 58945.5 to 58951.6). The coincidence between AT2020hur and activity window of FRB 180916B is interesting and is an independent confirmation of the FRB 180916B-AT2020hur association. Since the observation details of the MASTER-Kislovodsk telescope are unknown, this coincidence may be due to an observational selection effect. Another possibility is that the optical counterpart is subject to the same periodic modulation as the FRB 180916B. This may be due to the optical counterpart is directly related to the FRB emission, and the optical emission and FRB emission have similar Doppler beaming angle and direction, as implied by the prompt FRB multiwavelength counterparts (e.g. \citealt{Metzger2019,Yang2019b,Beloborodov2020}). In this case, each optical counterpart may correspond to an FRB. So we check the transit time of CHIME on 2020-04-08 and 2020-04-09, however, the exposure times of AT2020hur did not overlap with CHIME observations. There are not reported FRBs at the exposure times of AT2020hur. Interestingly, a total of nine low-frequency FRBs down to 120MHz were detected by LOFAR in the same activity window \citep{Pastor2020}, which have a time delay of $\sim1-3$ days with respect to AT2020hur. Whether there is a link between low-frequency FRB and optical counterpart needs to be confirmed by more observations.

\section{Conclusions} \label{sec:con}

In this paper, we perform a systematic search for ATs whose positions are consistent with FRBs. We find that one unclassified optical transient AT2020hur is spatially coincident with the repeating FRB 180916B. The chance possibility for the AT2020hur-FRB 180916B association is about 0.04\%, which corresponds to a significance of $3.5\sigma$. We first give a possible explanation for the AT2020hur-FRB 180916B association, in which FRB 180916B is powered by a flaring magnetar, while AT2020hur originates from the afterglow of one or more energetic GF. However, the derived isotropic kinetic energy of GF is too large compared to the typical GFs, and there is a lot of fine tuning and coincidences required for this specific model. Therefore, we conclude that the giant flare afterglow model is not a promising model to explain the AT2020hur. We also discuss other possible origins of AT2020hur. One possibility is that AT2020hur may consist of two or more optical flares originating from short-duration prompt FRB counterpart. These optical flares might be consistent with the theoretical predictions of the models of \cite{Metzger2019} and \cite{Beloborodov2020} or the model in which such flares originate from ``two-zone" IC scattering of FRB emission. Besides, the coincidence between AT2020hur and activity window of FRB 180916B is an independent confirmation of the FRB 180916B-AT2020hur association, and this coincidence may suggest that the optical counterpart is subject to the same periodic modulation as the FRB 180916B. If AT2020hur originates from the prompt FRB counterparts, future simultaneous detection of FRBs and their optical counterparts may reveal the physical origin of FRBs.

\acknowledgments
We are very grateful to an anonymous referee for helpful comments that have allowed us to improve our manuscript. We thank Hai-Ming Zhang, Qian-Cheng Liu, and Yuan-Pei Yang for helpful discussions. This work was supported by the National Key Research and Development Program of China (grant No. 2017YFA0402600), the National SKA Program of China (grant No. 2020SKA0120300), and the National Natural Science Foundation of China (grants No. 11833003 and U1831207). S.Q.Z. is supported by the China Post-doctoral Science Foundation (grant No. 2021TQ0325).

\vspace{5mm}
\software{emcee \citep{emcee}}

\clearpage

\appendix

\section{A List of 50 FRB-AT Pairs with Nearest Distance} \label{sec:pairs}

We list the 50 FRB-AT pairs with nearest distance in Table \ref{tab:pairs}.
\begin{deluxetable}{lllllllllllll}[h]\label{tab:pairs}
\tabletypesize{\tiny}
\tablewidth{0pt}
\tablecaption{50 FRB-AT Pairs with Nearest Distance}
\tablehead{
\colhead{FRB Name}	&	\colhead{FRB R.A.}	&	\colhead{FRB Dec.}	&	\colhead{FRB DM}	&	\colhead{AT Name}	&	\colhead{AT R.A.}	&	\colhead{AT Dec.}	&	\colhead{AT Type}	&	\colhead{AT Redshift}	&	\colhead{$\Delta t$}	&	\colhead{Distance}	&	\colhead{Chance Possibility}	\\
\colhead{ }	&	\colhead{(deg)}	&	\colhead{(deg)}	&	\colhead{(pc cm$^{-3}$)}	&	\colhead{ }	&	\colhead{(deg)}	&	\colhead{(deg)}	&	\colhead{ }	&	\colhead{ }	&	\colhead{(days)}	&	\colhead{(deg)}	&	\colhead{ }
}
\startdata
FRB 20180916B (rep)	&	$29.5031258$	&	$65.7167542$	&	347.8	&	AT2020hur	&	29.503125	&	65.71675	&	-	&	-	&	-570	&	0.0000042 	&	0.0418\%	\\
	&	$\pm0.0000006$	&	$\pm0.0000006$	&		&		&	$\pm0.0003$	&	$\pm0.0003$	&		&		&		&	(15 mas)	&	(3.5$\sigma$)	\\
FRB 20200405A	&	$198.36\pm1.50$	&	$36.59\pm1.50$	&	212.298	&	AT2019wur	&	198.3646667	&	36.5938	&	-	&	-	&	113	&	0.00019 	&	100\%	\\
FRB 20120127A	&	$348.78\pm0.13$	&	$-18.43\pm0.13$	&	553.3	&	AT2020abtj	&	348.7736792	&	-18.43970833	&	-	&	-	&	-3236	&	0.01 	&	100\%	\\
FRB 20190502A	&	$165.01\pm0.14$	&	$59.95\pm0.13$	&	626	&	SN2020adii	&	164.9727917	&	59.95431111	&	SN Ia	&	0.045077	&	-598	&	0.02 	&	100\%	\\
FRB 20190531E	&	$15.20\pm0.26$	&	$0.54\pm0.37$	&	328.4	&	SDSS-IISN17999	&	15.1956	&	0.517973	&	Ia	&	-	&	4277	&	0.02 	&	100\%	\\
FRB 20181027A	&	$131.90\pm0.22$	&	$-4.24\pm0.34$	&	726.3	&	AT2019jcp	&	131.8896917	&	-4.261511111	&	-	&	-	&	-92	&	0.02 	&	100\%	\\
FRB 20200405A	&	$198.36\pm1.50$	&	$36.59\pm1.50$	&	212.298	&	SN1985L	&	198.3407917	&	36.60916944	&	SN II	&	-	&	12715	&	0.02 	&	100\%	\\
FRB 20190131C	&	$166.45\pm0.22$	&	$10.43\pm0.27$	&	506.3	&	AT2017aih	&	166.4549833	&	10.40249444	&	-	&	-	&	734	&	0.03 	&	100\%	\\
FRB 20190330A	&	$204.09\pm0.22$	&	$38.36\pm0.09$	&	509.5	&	AT2017jg	&	204.0570917	&	38.34797778	&	-	&	-	&	810	&	0.03 	&	100\%	\\
FRB 20190330A	&	$204.09\pm0.22$	&	$38.36\pm0.09$	&	509.5	&	SN2019fck	&	204.0542292	&	38.35506111	&	SN Ia	&	0.02	&	-44	&	0.03 	&	100\%	\\
FRB 20181201B	&	$273.38\pm0.22$	&	$56.31\pm0.21$	&	875.1	&	AT2019pmr	&	273.3270417	&	56.32178333	&	-	&	-	&	-279	&	0.03 	&	100\%	\\
FRB 20190502B	&	$212.04\pm0.13$	&	$64.44\pm0.04$	&	917.1	&	SN2020iaf	&	212.0472292	&	64.40559167	&	SN Ia	&	0.11	&	-357	&	0.03 	&	100\%	\\
FRB 20190318A	&	$324.11\pm0.11$	&	$74.46\pm0.18$	&	420.5	&	AT2016ejk	&	324.0963958	&	74.41849167	&	-	&	-	&	964	&	0.04 	&	100\%	\\
FRB 20200405A	&	$198.36\pm1.50$	&	$36.59\pm1.50$	&	212.298	&	SN2001gd	&	198.3495417	&	36.63825	&	SN IIb	&	-	&	6707	&	0.05 	&	100\%	\\
FRB 20190226B	&	$273.57\pm0.23$	&	$61.81\pm0.25$	&	630.8	&	AT2021giu	&	273.6342292	&	61.85265556	&	-	&	-	&	-751	&	0.05 	&	100\%	\\
FRB 20190415A	&	$182.38\pm0.21$	&	$71.28\pm0.23$	&	632.4	&	AT2021yau	&	182.354	&	71.22713056	&	-	&	-	&	-873	&	0.05 	&	100\%	\\
FRB 20190415A	&	$182.38\pm0.21$	&	$71.28\pm0.23$	&	632.4	&	Gaia21edf	&	182.3540083	&	71.22712778	&	-	&	-	&	-873	&	0.05 	&	100\%	\\
FRB 20190604A (rep)	&	$218.78\pm0.16$	&	$53.28\pm0.17$	&	553.2	&	AT2020itk	&	218.6907542	&	53.27175556	&	-	&	-	&	-332	&	0.05 	&	100\%	\\
FRB 20181203C	&	$198.48\pm0.10$	&	$72.94\pm0.19$	&	2442.4	&	AT2021txl	&	198.3843125	&	72.98632222	&	-	&	-	&	-960	&	0.05 	&	100\%	\\
FRB 20190417A (rep)	&	$294.85\pm0.20$	&	$59.40\pm0.27$	&	1378.1	&	AT2020pnl	&	294.9457292	&	59.36925833	&	-	&	-	&	-450	&	0.06 	&	100\%	\\
FRB 20181216A	&	$306.28\pm0.22$	&	$53.53\pm0.23$	&	541.9	&	AT2021pyw	&	306.3755333	&	53.54190833	&	-	&	-	&	-910	&	0.06 	&	100\%	\\
FRB 20190417A (rep)	&	$294.85\pm0.20$	&	$59.40\pm0.27$	&	1378.1	&	SN2019pfb	&	294.9470917	&	59.36799722	&	SN Ib	&	0.03683	&	-137	&	0.06 	&	100\%	\\
FRB 20200405A	&	$198.36\pm1.50$	&	$36.59\pm1.50$	&	212.298	&	SN1950C	&	198.4375	&	36.585	&	-	&	0.0029	&	25529	&	0.06 	&	100\%	\\
FRB 20190601A	&	$190.11\pm0.21$	&	$62.72\pm0.21$	&	2228.9	&	AT2021nts	&	190.0332958	&	62.77015833	&	-	&	-	&	-728	&	0.06 	&	100\%	\\
FRB 20190628A	&	$199.06\pm0.22$	&	$51.75\pm0.22$	&	745.7	&	AT2020hvg	&	199.0706083	&	51.81364167	&	-	&	-	&	-237	&	0.06 	&	100\%	\\
FRB 20110703A	&	$352.71\pm0.13$	&	$-2.87\pm0.13$	&	1103.6	&	SN2005co	&	352.72325	&	-2.938444444	&	SN Ia	&	0.017	&	2204	&	0.07 	&	100\%	\\
FRB 20160206A	&	$15.25$	&	$41.63$	&	1262	&	AT2016irr	&	15.246975	&	41.56181944	&	-	&	-	&	-302	&	0.07 	&	-	\\
FRB 20190617C	&	$134.37\pm0.21$	&	$35.70\pm0.21$	&	637.3	&	AT2021eaz	&	134.4453042	&	35.66880556	&	-	&	-	&	-600	&	0.07 	&	100\%	\\
FRB 20151125A	&	$22.75$	&	$30.98$	&	273	&	AT2017auz	&	22.71595417	&	30.91736667	&	-	&	-	&	-449	&	0.07 	&	-	\\
FRB 20190404A	&	$143.10\pm0.10$	&	$36.90\pm0.17$	&	1353.8	&	AT2021zjb	&	143.1458458	&	36.8394	&	-	&	-	&	-903	&	0.07 	&	100\%	\\
FRB 20190117D	&	$208.87\pm0.14$	&	$31.68\pm0.20$	&	1175.9	&	AT2021put	&	208.8003042	&	31.64066111	&	-	&	-	&	-823	&	0.07 	&	100\%	\\
FRB 20190323A	&	$112.37\pm0.20$	&	$34.46\pm0.21$	&	855.7	&	AT2019cpd	&	112.3361458	&	34.52634444	&	-	&	-	&	-2	&	0.07 	&	100\%	\\
FRB 20181019B	&	$37.87\pm0.22$	&	$68.18\pm0.24$	&	723	&	AT2017fop	&	37.95333333	&	68.24846111	&	-	&	-	&	459	&	0.08 	&	100\%	\\
FRB 20190317E	&	$274.37\pm0.19$	&	$13.25\pm0.23$	&	800.7	&	AT2019umo	&	274.3573083	&	13.17541111	&	-	&	-	&	-212	&	0.08 	&	100\%	\\
FRB 20190604A (rep)	&	$218.78\pm0.16$	&	$53.28\pm0.17$	&	553.2	&	AT2017bex	&	218.6533	&	53.27466944	&	-	&	-	&	838	&	0.08 	&	100\%	\\
FRB 20190223B	&	$311.62\pm0.23$	&	$60.56\pm0.24$	&	535.4	&	AT2020adoq	&	311.661125	&	60.63415	&	-	&	-	&	-669	&	0.08 	&	100\%	\\
FRB 20171209A	&	$237.60\pm0.13$	&	$-46.17\pm0.13$	&	1457.4	&	GRB110715A	&	237.665	&	-46.237	&	LGRB	&	0.82	&	2339	&	0.08 	&	100\%	\\
FRB 20190907A (rep)	&	$122.21\pm0.18$	&	$46.27\pm0.23$	&	307.32	&	SN2011km	&	122.3036083	&	46.31355833	&	SN Ia	&	0.0459	&	3156	&	0.08 	&	100\%	\\
FRB 20181017D	&	$9.12$	&	$11.33$	&	1845.2	&	SN1996bl	&	9.074870833	&	11.39458056	&	SN Ia	&	-	&	8041	&	0.08 	&	-	\\
FRB 20181213A	&	$127.66\pm0.20$	&	$73.87\pm0.10$	&	677.7	&	AT2017buw	&	127.4998167	&	73.93592778	&	-	&	-	&	651	&	0.08 	&	100\%	\\
FRB 20181223C	&	$181.05\pm0.19$	&	$27.58\pm0.20$	&	111.6	&	AT2021pgz	&	181.1395708	&	27.59588611	&	-	&	-	&	-898	&	0.08 	&	100\%	\\
FRB 20181027A	&	$131.90\pm0.22$	&	$-4.24\pm0.34$	&	726.3	&	AT2020aeca	&	131.9806208	&	-4.229866667	&	-	&	-	&	-775	&	0.08 	&	100\%	\\
FRB 20190608A	&	$359.23\pm0.22$	&	$19.17\pm0.24$	&	719.8	&	AT2021aux	&	359.2160875	&	19.25164722	&	-	&	-	&	-589	&	0.08 	&	100\%	\\
FRB 20190417B	&	$174.87\pm0.20$	&	$64.72\pm0.21$	&	1151.7	&	SN2017hkz	&	174.84675	&	64.80435	&	SN Ia	&	0.0435	&	548	&	0.08 	&	100\%	\\
FRB 20171004A	&	$179.40\pm0.20$	&	$-11.90\pm0.17$	&	304	&	AT2021eft	&	179.4469167	&	-11.97328333	&	-	&	-	&	-1226	&	0.09 	&	100\%	\\
FRB 20150610A	&	$161.11\pm0.13$	&	$-40.09\pm0.13$	&	1593.9	&	Gaia15acl	&	160.9976667	&	-40.12052778	&	-	&	-	&	108	&	0.09 	&	100\%	\\
FRB 20190125B	&	$231.45\pm0.25$	&	$50.54\pm0.23$	&	177.9	&	1993U	&	231.3122917	&	50.57741944	&	QSO	&	-	&	9351	&	0.10 	&	100\%	\\
FRB 20190621B	&	$193.14\pm0.23$	&	$55.64\pm0.23$	&	1059.5	&	AT2019jwr	&	193.3057958	&	55.66150833	&	-	&	-	&	13	&	0.10 	&	100\%	\\
FRB 20181017D	&	$9.12$	&	$11.33$	&	1845.2	&	PS15bvm	&	9.117583333	&	11.23697222	&	-	&	-	&	1138	&	0.10 	&	-	\\
FRB 20190531E	&	$15.20\pm0.26$	&	$0.54\pm0.37$	&	328.4	&	SDSS-IISN15565	&	15.113	&	0.589886111	&	SN II	&	-	&	4659	&	0.10 	&	100\%	\\
\enddata
\tablecomments{The chance posibility $P$ can be calculated as $P=1-(1-P_{1})^{609}$, where $P_{1}=1-\exp\{-112915[1-\cos (D+\delta_{\rm{FRB}}+\delta_{\rm{AT}})]/2\}$.  Here 609 is the total number of FRBs, 112915 is the total number of ATs, $D$ is the distance between a FRB and a AT, $\delta_{\rm{FRB}}=\sqrt{\delta_{\rm{FRB,RA}}\delta_{\rm{FRB,Dec}}}$ is the error radius of a FRB, and $\delta_{\rm{AT}}$ is the error radius of a AT with a typical value $\delta_{\rm{AT}}\sim1$ arcsec.}
\end{deluxetable}

\clearpage

\section{Fitting Results from the Off-axis Configuration} \label{sec:multi2}

\begin{figure}[h]
\centering
\includegraphics[width=0.4\textwidth,angle=0]{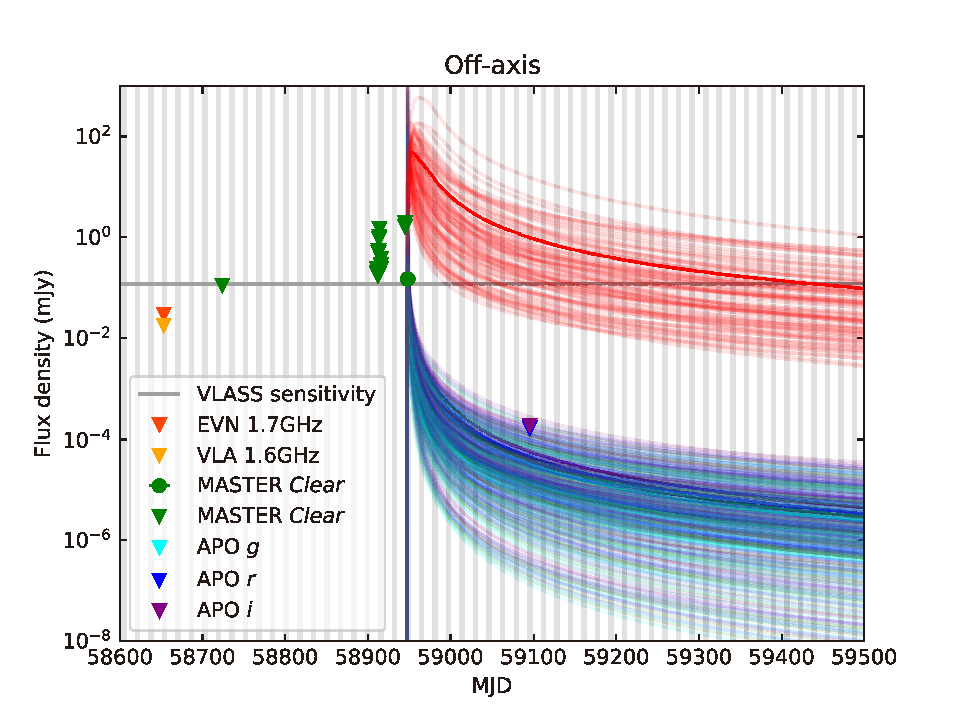}
\includegraphics[width=0.4\textwidth,angle=0]{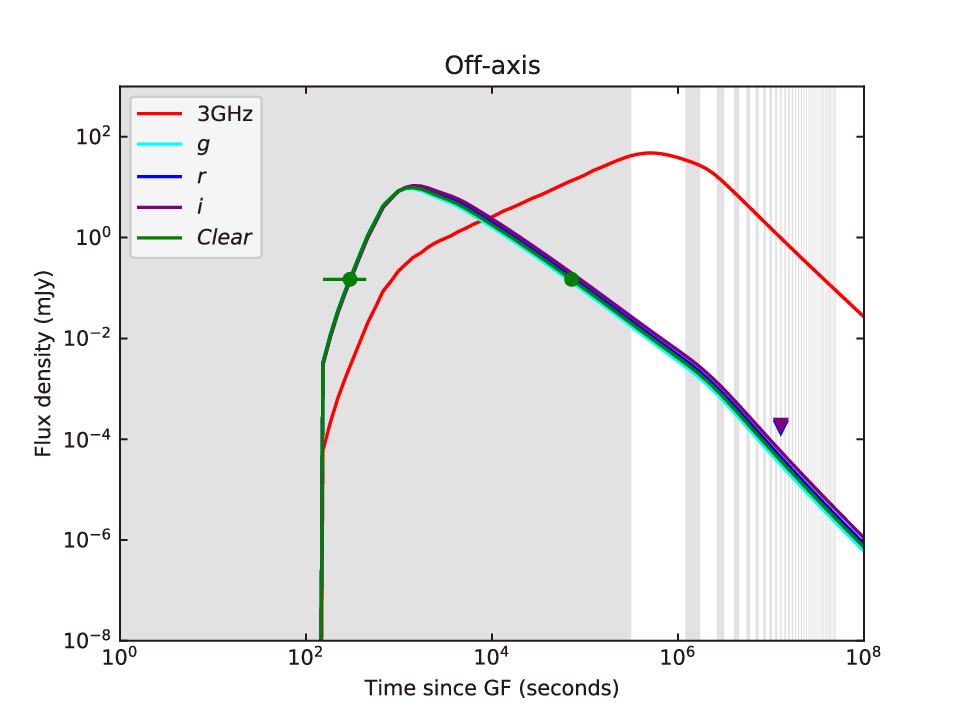}
\includegraphics[width=0.4\textwidth,angle=0]{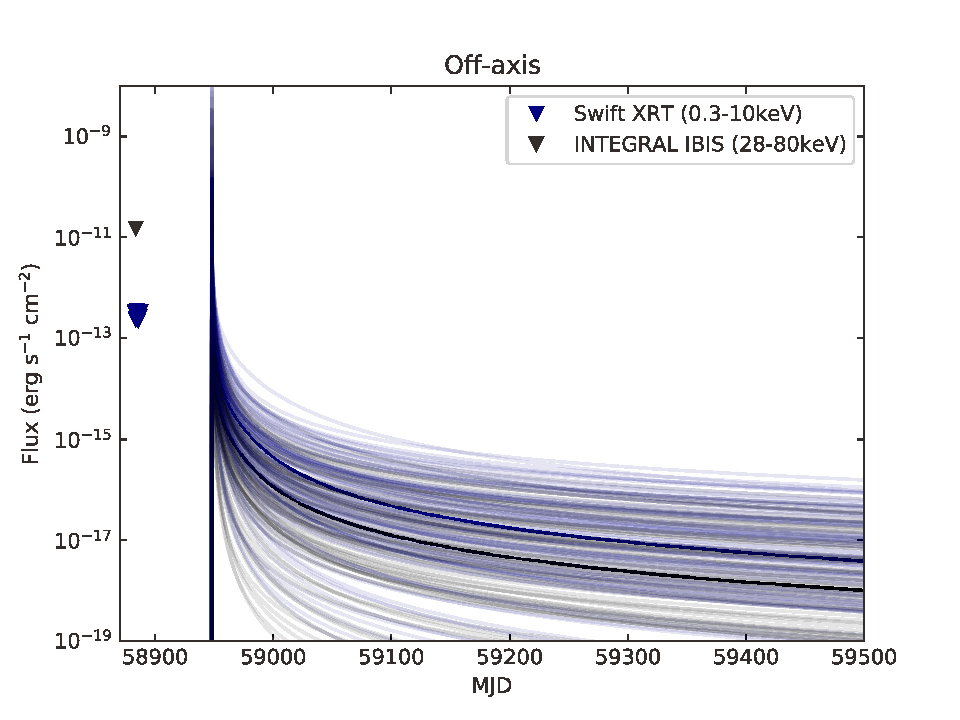}
\includegraphics[width=0.4\textwidth,angle=0]{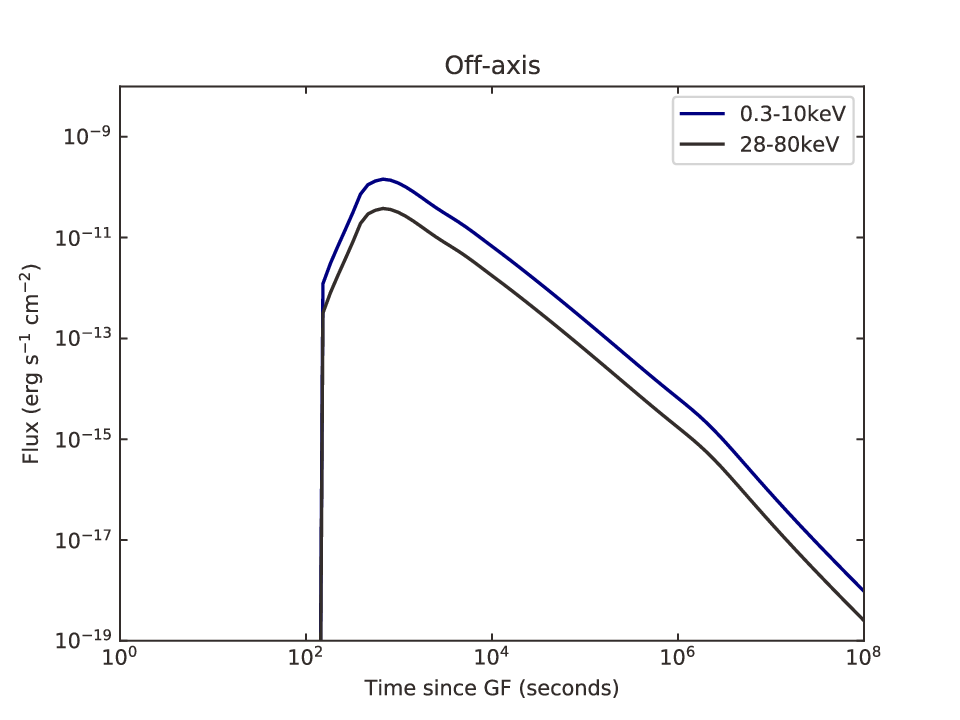}
\caption{The standard forward shock afterglow model fit to AT2020hur and multiwavelength observations of FRB 180916B. Upper left: the optical and radio lightcurves in a linear time scale. Green circles are the $C$-band observations of AT2020hur \citep{Lipunov2020}. Green upside down triangles are the $C$-band upper limits of FRB 180916B \citep{Zhirkov2020}. Cyan, blue and purple upside down triangles are the $g$-, $r$-, and $i$-band upper limits of FRB 180916B observed by Apache Point Observatory (APO; \citealt{Kilpatrick2021}). Orange and yellow upside down triangles are the upper limits on the persistent radio emission associated with FRB 180916B observed by the European Very-long-baseline-interferometry Network (EVN) and the Karl G. Jansky Very Large Array (VLA; \citealt{Marcote2020}). Gray horizontal line is the single epoch sensitivity of the 3 GHz VLA Sky Survey (VLASS; \citealt{Lacy2020}). Upper right: the optical and radio lightcurves in a logarithmic times cale. Lower left: the X-ray and $\gamma$-ray lightcurves in a linear time scale. Navy upside down triangles are the X-ray upper limits observed by Swift XRT \citep{Tavani2020a}. Black upside down triangle is the $\gamma$-ray upper limits observed by INTEGRAL IBIS \citep{Panessa2020}. Lower right: X-ray and $\gamma$-ray lightcurves in a logarithmic time scale. The gray shaded regions in the upper left and right panels correspond to the predicted activity days of FRB 180916B for a period of 16.29 days and a 6.1 day activity windows \citep{Pastor2020}
\label{fig:multi2}}
\end{figure}

We also show the fitting results from the off-axis configuration in Figure \ref{fig:multi2}.

\clearpage

\section{Corner Plots} \label{sec:corner}

\begin{figure}[h]
\centering
\includegraphics[width=0.55\textwidth,angle=0]{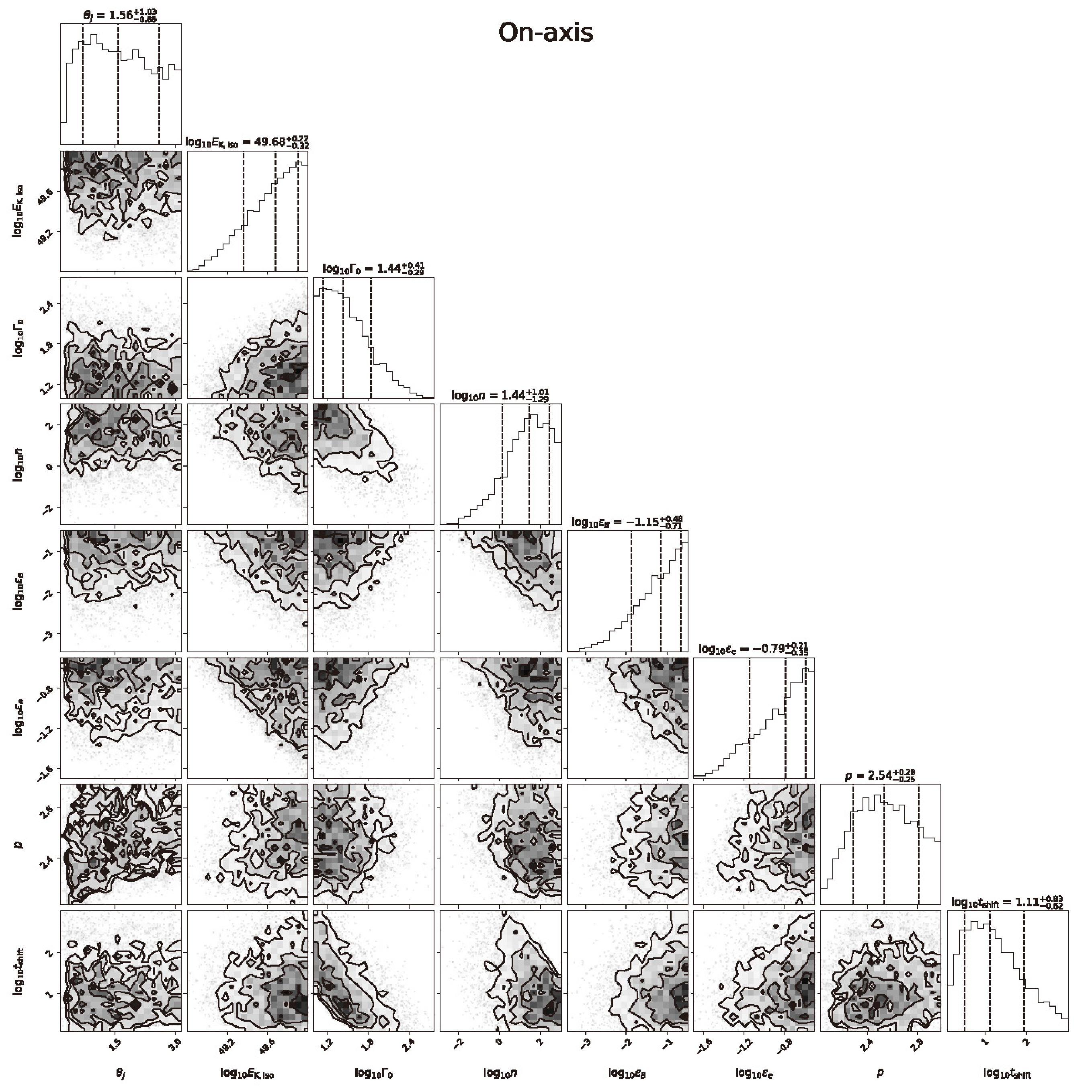}
\includegraphics[width=0.55\textwidth,angle=0]{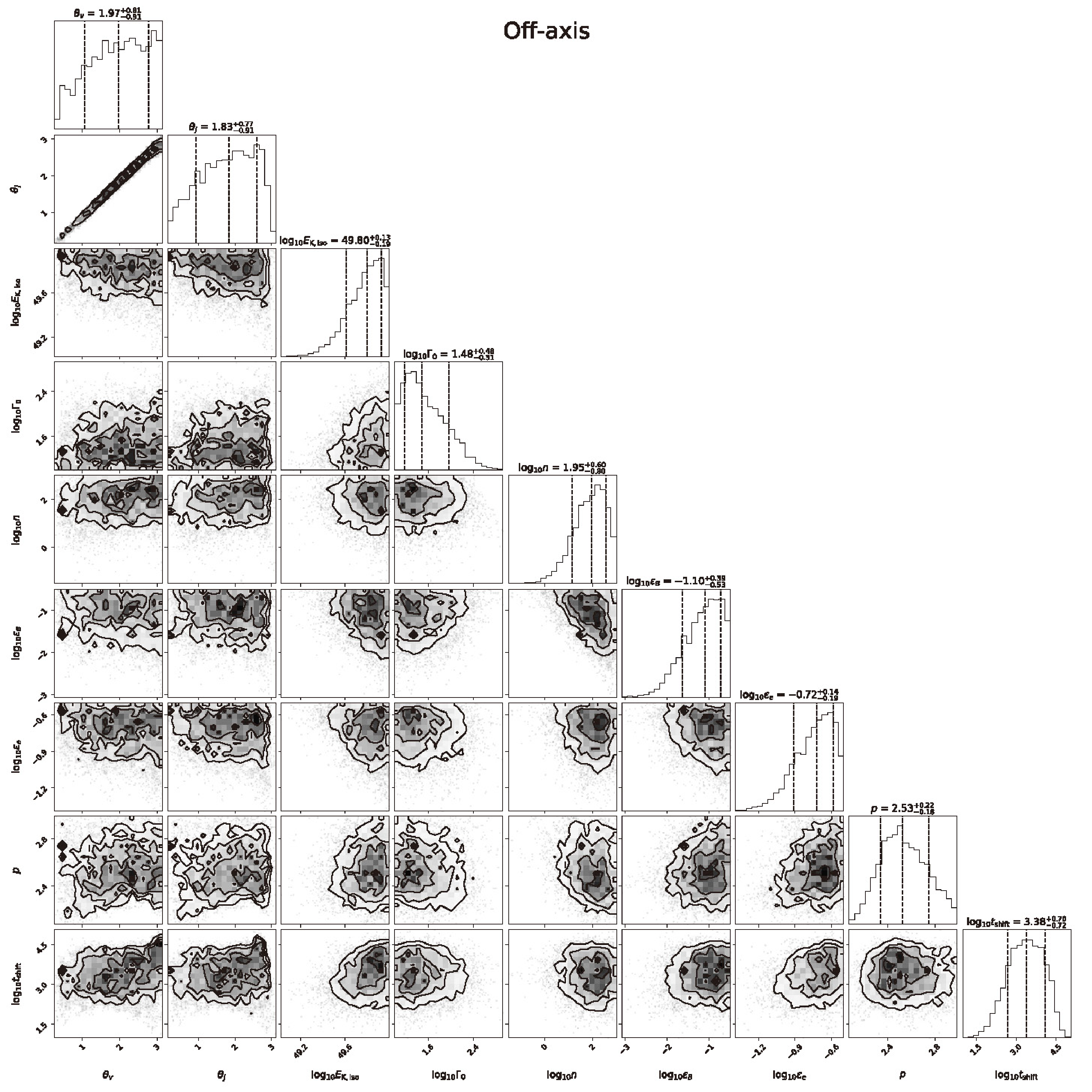}
\caption{Corner plots showing the one and two dimensional projections of the posterior probability distributions of parameters for the standard forward shock afterglow model with on-axis and off-axis configurations, respectively.
\label{fig:corner}}
\end{figure}

We show the corner plots for the standard forward shock afterglow model with on-axis and off-axis configurations in Figure \ref{fig:corner}.

\clearpage


\end{document}